\documentclass[]{emulateapj}
\usepackage{graphicx,natbib,color,setspace,amssymb, epsfig,textcomp,indentfirst,verbatim,amsmath}

\slugcomment{Version of \today}
\newcommand*\al{$\alpha$}
\newcommand*\ad{$\alpha$-}
\newcommand*\diff{\mathop{}\!\mathrm{d}}

\shorttitle{Ages of APOGEE Giants}
\shortauthors{Feuillet et al.}

\begin{document}
\label{firstpage}

\title{Determining Ages of APOGEE Giants with Known Distances}

\author{Diane~K.~Feuillet\altaffilmark{1},
Jo~Bovy\altaffilmark{2},
Jon~Holtzman\altaffilmark{1},
L\'eo~Girardi\altaffilmark{3}$^,$\altaffilmark{4},
Nick~MacDonald\altaffilmark{5},
Steven~R.~Majewski \altaffilmark{6},
David~L.~Nidever\altaffilmark{7}$^,$\altaffilmark{8}
}

\altaffiltext{1}{Department of Astronomy, New Mexico State University, Las Cruces, NM 88003, USA; feuilldk@nmsu.edu}
\altaffiltext{2}{Department of Astronomy and Astrophysics, University of Toronto, 50 St. George Street, Toronto, ON, M5S 3H4, Canada}
\altaffiltext{3}{Osservatorio Astronomico di Padova $-$ INAF, Vicolo dell'Osservatorio 5, I-35122 Padova, Italy}
\altaffiltext{4}{Laborat\'orio Interinstitucional de e-Astronomia $-$ LIneA, Rua Gal. Jos\'e Cristino 77, Rio de Janeiro, RJ $-$ 20921-400, Brazil}
\altaffiltext{5}{Department of Astronomy, University of Washington, Seattle, WA 98195, USA}
\altaffiltext{6}{Department of Astronomy, University of Virginia, Charlottesville, VA 22904}
\altaffiltext{7}{Large Synoptic Survey Telescope, 950 North Cherry Ave, Tucson, AZ 85719}
\altaffiltext{8}{Steward Observatory, 933 North Cherry Ave, Tucson, AZ 85719}

\begin{abstract}

We present a sample of 705 local (d $<400$ pc) red giant stars observed using the New Mexico State University 1~m telescope with the Sloan Digital Sky Survey III Apache Point Observatory Galactic Evolution Experiment (APOGEE) spectrograph, for which we estimate stellar ages and the age distribution from the high-resolution spectroscopic stellar parameters and accurate distance measurements from Hipparcos. The high-resolution (R $\sim$ 23,000), near infrared ($H$-band, 1.5-1.7~\micron) APOGEE spectra provide measurements of the stellar atmospheric parameters (temperature, surface gravity, [M/H], and [$\alpha$/M]). Due to the smaller uncertainties in surface gravity possible with high-resolution spectra and accurate Hipparcos distance measurements, we are able to calculate the stellar masses to within 40\%. For red giants, the relatively rapid evolution of stars up the red giant branch allows the age to be constrained based on the mass. We examine methods of estimating age using both the mass-age relation directly and a Bayesian isochrone matching of measured parameters, assuming a constant star formation history (SFH). To improve the prior on the SFH, we use a hierarchical modeling approach to constrain the parameters of a model SFH from the age probability distribution functions of the data. The results of an \ad dependent Gaussian SFH model shows a clear relation between age and [\al/M] at all ages. Using this SFH model as the prior for an empirical Bayesian analysis, we construct a full age probability distribution function and determine ages for individual stars. The age-metallicity relation is flat, with a slight decrease in [M/H] at the oldest ages and a $\sim$ 0.5~dex spread in metallicity. For stars with ages $\lesssim 1$~Gyr we find a smaller spread, consistent with radial migration having a smaller effect on these young stars than on the older stars. This method of estimating ages of red giants is developed with the intent of estimating ages for the much larger sample of APOGEE survey giants that will have parallax measurements from Gaia.
\end{abstract}

\keywords{Galaxy: disk - Galaxy: evolution - stars: abundances - stars: fundamental parameters}

\section{Introduction}

As the only galaxy for which detailed measurements of large numbers of individual stars are currently possible, the Milky Way serves as a crucial testing ground for exploring galactic evolution through detailed stellar observations. The chemical composition of stars reflects the composition of the interstellar medium from which they formed and contains information about the galactic gas history, tracing the evolution of stellar populations, the merger history, and the star formation history (SFH). Each stellar population in a galaxy contains a snapshot of the gas at the time of its birth. In a closed system the overall metallicity of the gas increases with time as metals are created in stars and recycled back to form new stars. Galaxies are not closed systems, but have gas inflow and outflow, and are not enriched homogeneously. The interstellar medium is not uniform across the Galactic disk; enriched material is pulled towards the central regions, while pristine gas accretes onto the outer regions. Much work has been done to include these influences in model of Galactic evolution \citep{Matteucci1989, Chiappini1997, Chang1999, Chiappini2001, Schonrich2009a, Minchev2013}. Radial mixing of stars creates an additional complication, moving stars from their birth radii and potentially diluting any radial metallicity gradients \citep{Wielen1996, Sellwood2002, Roskar2008, Loebman2011, Hayden2015}. Detailed chemical abundance measurements for large samples of stars can help identify the different stellar populations of our Galaxy and understand the Galactic enrichment history.

Abundance measurements for large samples of stars throughout the Galaxy allow us to explore the chemical evolution of the stellar populations \citep{Lee2011, Boeche2013, Hayden2014, Nidever2014, Hayden2015}. The evolution of the stellar populations is often characterized by the overall metallicity ([Fe/H]), however, the rate of metal enrichment depends on the local star formation rate. The various nucleosynthetic processes by which metals are created also enrich families of elements on different timescales \citep{Tinsley1979, Matteucci1986}. This makes it difficult to place an absolute timescale on Galactic evolution. In addition, enrichment may not be uniform across the Galactic disk, complicating direct comparisons of the chemical history of the inner and outer disks using metallicity alone. In order to directly compare the abundance evolution of all Galactic stellar populations, an absolute timescale is needed. Therefore, we need to measure absolute ages for large samples of stars.

\citet{Soderblom2010} discusses the difficulties of age determination methods for field stars, showing that stellar ages cannot be directly measured but must be estimated from models or empirical methods. Although ages of subgiants can be found to within 1~Gyr through isochrone matching \citep[see e.g. ][]{Haywood2013, Bensby2014}, the atmospheric parameters of dwarfs and giants (temperature, surface gravity, and metallicity) are fairly degenerate with age. 
In this paper, we present a method for determining absolute ages of field red giant stars with accurately measured distances and high resolution spectroscopic atmospheric parameters. This method takes advantage of the short duration of the giant branch stage to apply a mass-age relation to the giant branch and estimate ages.

Recently, spectroscopic surveys have made great progress in obtaining spectra of large samples of stars across the Galaxy. 
SEGUE \citep{Yanny2009,Lee2011} and RAVE \citep{Kordopatis2013} have provided low resolution spectra for over 400,000 stars. However, higher resolution spectra are needed to accurately measure atmospheric parameters and multi-element abundances of large samples of stars. The Gaia-ESO Survey \citep[GES,][]{Gilmore2012, Randich2013} is a high resolution survey on the VLT using the GIRAFFE and UVES spectrographs to measure abundances of up to 24 individual elements of 100,000 stars. The Galactic Archaeology with HERMES (GALAH) survey \citep{Freeman2012, Zucker2012} is a high resolution spectroscopic survey on the 4m Anglo-Australian Telescope and will provide abundances of over 20 elements for $10^6$ stars in the Gaia footprint. 

Although these surveys continue to observe stars at greater distances from the Sun, by working at optical wavelengths, they are still extremely limited by the dust in the plane of the disk. The Sloan Digital Sky Survey (SDSS) III Apache Point Observatory Galactic Evolution Experiment \citep[APOGEE,][]{Eisenstein2011} is a high-resolution, near-infrared (NIR), spectroscopic survey of red giant stars in the Milky Way. With the goal of studying the evolution of the Milky Way, APOGEE samples all of the stellar components in the Galaxy with a particular focus on the inner disk and bulge. The APOGEE spectrograph is uniquely designed to study this region of the Galaxy because the obscuring effects of dust in the plane of the disk are about six times lower in the NIR than in the optical. The SDSS Data Release 12 \citep[DR12, ][]{Alam2015} contains 146,000 APOGEE stars and the SDSS IV extension, APOGEE2, will expand the sample to over 500,000 stars observed from both the Northern and Southern hemispheres, allowing for complete Galactic coverage. As part of DR12, the APOGEE instrument was used with the New Mexico State University (NMSU) 1~m telescope to observe a sample of nearby red giant stars. 

Section \ref{obs} describes the observations and the sample, with a brief description of the connection of the APOGEE instrument to the NMSU 1~m telescope. A description of the analysis of the spectra to determine atmospheric parameters, elemental abundances, and ages is presented in Sections \ref{aspcap} and \ref{ages}. In Section \ref{agetrends} we discuss our results, finding a clear relation between [\al/M] and age, lack of a strong age-metallicity relation, and increasing velocity dispersion with age. We summarize our conclusions in Section \ref{conclusion}.

\section{Observations and Sample Selection}
\label{obs}

\subsection{1m+APOGEE}
\label{1m+A}

The APOGEE instrument is a high resolution (R $\sim$ 23,000), fiber fed, near infrared ($H$-band, 1.51 - 1.7~\micron) spectrograph (Wilson, in prep) and is usually used with plug plates that feed 300 spectra from the SDSS 2.5~m telescope \citep{Gunn2006}. To observe single bright and nearby objects using the multi-object observing scheme of the Sloan 2.5~m telescope is an inefficient use of that valuable telescope time. Thus, 10 fibers were installed to connect the APOGEE instrument to the NMSU 1~m telescope about 50 yards away at the Apache Point Observatory (APO). The new fibers run underground from the APOGEE instrument port at the base of the 2.5~m to the NA2 port of the 1~m, where they are positioned at the focal plane of the NA2 port in a fixed linear configuration. This configuration allows for one science target fiber and nine sky fibers per observation. 

An advantage of this new observing capability, hereafter 1m+APOGEE, is the extended use of the APOGEE instrument with minimal additional operational cost. The NMSU 1~m telescope is operated robotically during science observations as described by \citet{Holtzman2010}, dramatically reducing the human-hours needed to complete these observations. The 1m+APOGEE can observe 20 - 40 stars per night, depending on the target magnitude. Targets observed with the 1m+APOGEE are typically limited to $H < 8$ because fainter targets require exposure times over an hour and are better suited for 2.5~m observations, and $H > 0$ because such bright targets saturate the 1~m guide camera as well as the APOGEE instrument.

APOGEE spectra observed with the 1~m telescope have been found to be comparable to those taken with the 2.5~m telescope, and have been used to help calibrate APOGEE survey data. A detailed description of these comparisons and calibration can be found in \citet{Holtzman2015}.

\subsection{Hipparcos Sample}
\label{hip}

As mentioned above, stellar age is a difficult parameter to determine, especially for red giant stars, which have similar atmospheric parameters across many ages. However, the age of a red giant star is determined by its mass. The mass of a star can be calculated from its luminosity, effective temperature, and surface gravity (see Section \ref{mass}). The APOGEE instrument allows for effective temperature and surface gravity measurements to within 92~K and 0.11~dex, respectively, for spectra with S/N $\approx 100$ \citep{Holtzman2015}. Luminosities can be calculated for nearby stars with parallax distances, such as those measured by the Hipparcos mission. Stellar ages could be estimated for a sample of nearby red giant stars observed with APOGEE for which distance measurements are available. The 1m+APOGEE capability is well designed to observe such a sample of bright nearby red giants that could also be directly compared to the larger, more distant APOGEE survey observations. 

\begin{figure}[t!]
\centering
\includegraphics[angle=90,width=0.45 \textwidth]{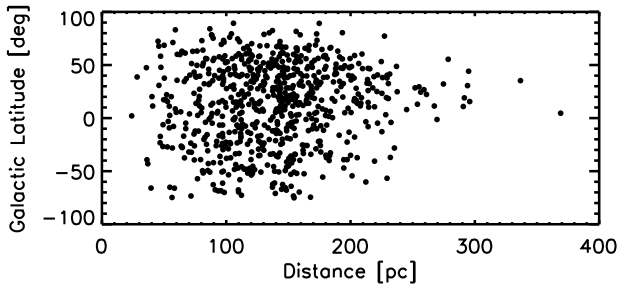}
\caption{ The distribution of the Hipparcos 1m+APOGEE red giant subsample used in this paper in distance and Galactic latitude.}
\label{spacedist}
\end{figure}

Observations of 750 red giant stars were made between October, 2012 and April, 2014 using the 1m+APOGEE. The sample was selected from the Hipparcos Catalog \citep{vanLeeuwen2007} to have parallaxes measured to within 10\%. A color cut of $(J-K) > 0.5$  and an absolute $H$ magnitude cut of $M_H < 2$ were used to select giants. Over 3500 targets meeting these criteria had a declination above $-20$, making them visible from APO. Observations were made of targets randomly selected each night from this final list. The observed sample is within 400 pc of the Sun covering all Galactic latitudes; it is displayed in Figure \ref{spacedist}. These data are included in the SDSS DR12 \citep{Alam2015}.

\section{Stellar Atmospheric Parameters}
\label{aspcap}

The spectra taken with the 1m+APOGEE are reduced and analyzed with the existing software used by the main APOGEE survey \citep{Nidever2015}. Similar to 2.5~m observations, the 1m+APOGEE observations have sky fibers positioned close to the science object on the sky for subtraction of the sky emission features. However, a different method was needed to treat the telluric absorption because no hot, relatively featureless star could be observed simultaneously, as is done for the multi-object observations. Instead, the atmospheric model spectrum is combined with a spectral template that best fits the target and adjusted to fit the telluric features in the observed target spectrum. This process is iterated to produce the telluric absorption spectrum that best matches the observed spectrum. 

Stellar parameters and abundances were derived from the spectra using the APOGEE Stellar Parameters and Chemical Abundances Pipeline \citep[ASPCAP,][]{Garcia2015}, which finds the best matching synthetic spectra to the observed spectra through $\chi^2$ minimization. For DR12, ASPCAP uses a six dimensional library of synthetic spectra \citep{Zamora2015} to simultaneously fit for effective temperature ($T_{\mbox{\small eff}}$), surface gravity ($\log g$), overall metallicity ([M/H]), and relative abundances of \ad elements ([\al/M]), carbon ([C/M]), and nitrogen ([N/M]). Molecular features from OH, CO, and CN have a large effect on the NIR spectrum, therefore \al, C, and N abundances must be included in the fit to atmospheric parameters. The microturbulence value is set using a linear relation to surface gravity derived from an analysis of a calibration subsample. The DR12 version of ASPCAP also includes a constant macroturbulence value of 6~km~s$^{-1}$. The DR12 grid of synthetic spectra covers $3000~\,~\mbox{K} < T_{\mbox{\small eff}} < 8000~\,~\mbox{K}$ and $0 < \log g < 5$. We refer the reader to \citet{Holtzman2015} for a characterization and calibration of the parameters. 

\citet{Holtzman2015} use several techniques to both internally and externally calibrate the raw ASPCAP results into agreement with the literature. We use these calibrated ASPCAP values in our analysis. The empirical uncertainty in $T_{\mbox{\small eff}}$ is 91.5~K  as determined through comparison of raw ASPCAP temperature values to photometric temperatures derived using the method described by \citet{GonzalezHernandez2009}. The empirical uncertainty in $\log g$ is 0.11~dex as determined through comparisons of targets in the Kepler field to asteroseismic data. Using internal scatter in [M/H] of cluster stars, the uncertainty in [M/H] is 0.05~dex. The ASPCAP [M/H] is externally calibrated using literature [Fe/H] for these clusters. ASPCAP does determine an [Fe/H] abundance separately from [M/H]. The differences between raw ASPCAP [M/H] and [Fe/H] abundances is small. Internal calibrations are applied to the [Fe/H] parameter using the same cluster method used for the [M/H] parameter. However, the [Fe/H] parameter is not externally calibrated because literature cluster abundances are not available for all 15 elements. In addition the [Fe/H] abundance is often used to convert [X/H] abundance into [X/Fe] abundances. Using an externally calibrated [Fe/H] abundance with uncalibrated [X/H] abundances could introduce systemic errors into elemental abundance ratios. For the purposes of this paper, the calibrated [M/H] parameter is comparable to the PARSEC isochrone [Fe/H] parameter \citep{Bressan2012}. Calibrations to raw ASPCAP parameters are only applied to giant stars. Although the sample presented in this paper was selected to contain giants based on color and magnitude, 39 of our stars are outside the ASPCAP calibration range and two stars have a calibrated $\log g > 3.8$ and were therefore not included in our analysis. There are also four stars with an ASPCAP $T_{\mbox{\small eff}}$ much hotter than the isochrones points with $(J-K) > 0.5$. Our final sample contains 705 stars.

As mentioned above, calibrations have been applied to the raw ASPCAP $\log g$ based on comparisons to targets with Kepler asteroseismic data. The comparisons to asteroseismic surface gravities showed a systematic offset in the ASPCAP $\log g$ between red giant branch (RGB) and red clump (RC) stars. The raw ASPCAP $\log g$ for RC stars shows a larger disagreement with asteroseismic $\log g$ than the RGB stars. The ASPCAP $\log g$ calibration is fit to the RGB locus because it is possible to identify the RC stars based on $T_{\mbox{\small eff}}$ and $\log g$ and apply an additional calibration. This additional calibration of RC stars is not done in DR12, however, the APOGEE RC catalogue \citep{Bovy2014a} does provide identification and $\log g$ correction of RC stars in the DR12 sample. 

In this paper we identify RC stars following the methods of \citet{Bovy2014a}. We apply an average correction to the raw ASPCAP $\log g$ derived from comparisons to high quality asteroseismic data, as described in Section 5.3 of \citet{Alam2015} and then identify RC stars using this average corrected $\log g$. The selection criteria for RC identification is taken from \citet{Bovy2014a} Equations (2), (3), (8), and (9). Because RC stars separate from the RGB more in absolute magnitude than in surface gravity, we include an addition cut on absolute Tycho $V$-band magnitude ($M_{V_T}$) as a function of $T_{\mbox{\small eff}}$:
\begin{eqnarray}
&&\frac{-T_{\mbox{\small eff}}}{600}  + 8.25 < M_{V_T}< 2.0 . \nonumber
\end{eqnarray}
This cut was chosen to remove stars brighter than the RC overdensity seen in our sample in $M_{V_T}$ vs $T_{\mbox{\small eff}}$ space. It is in agreement with the location of the RC evolutionary stage of the PARSEC isochrones. Using this criteria we find 324 RC stars in this sample. Once the RC stars were selected, the following $\log g$ correction was applied to the raw ASPCAP $\log g$ of just the RC stars.
\begin{equation}
\log g_{RC} = 
\begin{cases}
\log g_{raw} - 0.255 & \log g_{raw} < 1.0 \\
0.958 \log g_{raw} - 0.213 & 1.0 < \log g_{raw} < 3.8 \\
\log g - 0.3726 & 3.8 < \log g \nonumber
\end{cases}
\end{equation}
The DR12 corrected ASPCAP $\log g$ value is used for stars not identified as RC stars. Both RC and RGB stars are included in our analysis and are collectively referred to as giants.

\begin{figure}[t!]
\centering
\includegraphics[angle=90,width=0.43 \textwidth]{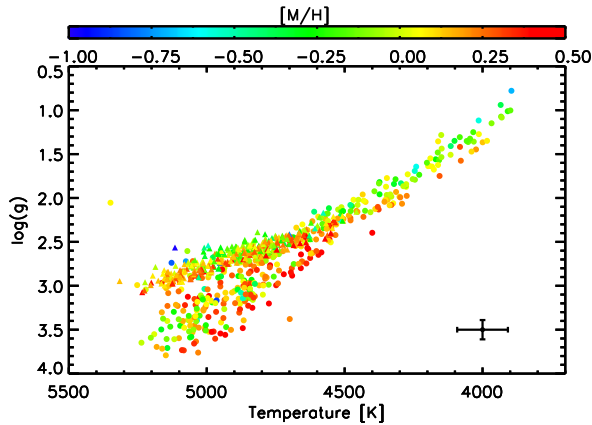}
\caption{ The HR diagram of this sample. The colors represent metallicity ([M/H]). The RGB stars are plotted as points and the RC stars, as identified in Section \ref{aspcap}, are plotted as triangles. The RC stars lie above the RGB sequence with lower $\log g$ at a given $T_{\mbox{\small eff}}$ for $5300 < T_{\mbox{\small eff}} < 4500.$ All stars shown are included in the age analysis.}
\label{HRD}
\end{figure}

\begin{figure}[t]
\centering
\includegraphics[angle=90,width=0.43 \textwidth]{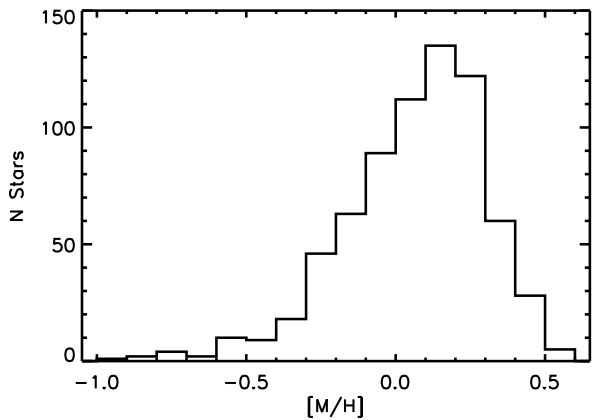}
\caption{ The metallicity distribution of our sample binned by 0.1~dex. The metallicity distribution of this sample agrees with literature results for the solar neighborhood, peaking just above solar metallicity with very few metal poor stars.}
\label{fehist}
\end{figure}

Figure \ref{HRD} shows the Hertzsprung-Russell (HR) diagram of the sample, where the color represents metallicity. Our samples covers the full giant branch, and includes 324 RC stars, as identified using the criteria above. The RC stars are indicated by triangles. We note that our metallicity range is $-1.0 <$ [M/H] $< 0.6$ with a peak just above solar, shown in Figure \ref{fehist}, which is expected for the solar neighborhood \citep[e.g. ][]{Adibekyan2012, Hayden2015}.

\section{Mass and Age}
\label{ages}
\subsection{Mass}
\label{mass}
While the ages of main sequence turn-off (MSTO) stars and subgiants can be accurately determined through comparisons of measured parameters (e.g. temperature, surface gravity, luminosity) to model isochrones \citep[e.g. ][]{Jorgensen2005, Holmberg2009, Haywood2013}, the measured properties of stars doing more advanced stages of evolution, such as RGB and RC stars, are very similar for stars of different ages and metallicities. Therefore finding ages of giants from isochrone matching using these parameters is not as precise as it is for subgiants or MSTO stars. The giant stage is brief compared to the main sequence lifetime of a star and the later is determined primarily by the initial mass and metallicity of the star. Thus, the age of a giant branch star can be constrained if the mass is known. We use stellar atmospheric parameters derived from high-resolution spectroscopy and accurately measured distances to calculate masses and estimate ages for local giant stars.

The mass of a star can be calculated from the photometrically and spectroscopically measured parameters using the following well-known relations:
\begin{equation}
\label{lum}
L = 4 \pi R^2 \sigma T_{\mbox{\small eff}}^4
\end{equation}
\begin{equation}
\label{g}
g = \frac{G M}{R^2}
\end{equation}
where $\sigma$ is the Stefan Boltzman constant, and $G$ is the gravitational constant.
For this sample, the luminosities were calculated from apparent Tycho $V$-band magnitudes ($V_T$), Hipparcos distances \citep{vanLeeuwen2007}, and model-based bolometric corrections (BCs). Due to the proximity of this sample to the Sun, extinction was assumed to be negligible. The BCs were determined using a 6th order polynomial fit to the PARSEC isochrone BC as a function of temperature. The high order of the fit was chosen to minimize the residuals to the fit across a large temperature range, resulting in a BC uncertainty of 0.03~mag. The solar bolometric magnitude and luminosity were taken from \citet{Bahcall1995} as  $M_{Bol,\odot} = 4.77$~mag and $L_{\odot} = 3.844 \times 10^{33}$~erg~s$^{-1}$. 
These stellar luminosities were applied to Equation (\ref{lum}) along with the spectroscopic temperatures to calculate radii. The mass of each star was then calculated from the radius and the spectroscopic surface gravity using Equation (\ref{g}). 

\begin{figure}[t!]
\centering
\includegraphics[angle=90,width=0.45 \textwidth]{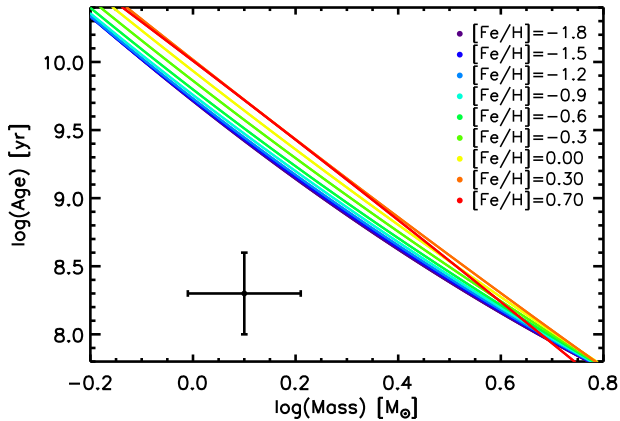}
\caption{ Each line represents the isochrone derived mass-age relation for a representative sample of metallicities. The color represents metallicity as indicated. The mass-age relation was fit to the PARSEC isochrone points of a single metallicity with a quadratic function in logarithmic space. Representative error bars are shown.}
\label{massagefit}
\end{figure}

To calculate the mass error, we adopt the APOGEE reported errors for the spectroscopically derived parameters. The $V_T$ magnitudes were taken from \citet{Anderson2012} and have a typical uncertainty of 0.05~mag. The sample was selected to have distance errors $< 10 \%$ in order to minimize the error contribution from the distance measurement. Due to the small errors in photometry and distance, the main source of error in the mass calculation comes from the ASPCAP uncertainty in surface gravity. This leads to an error in mass of 0.11~dex, about 30 \%.

\begin{figure}[t!]
\centering
\includegraphics[angle=90,width=0.45 \textwidth]{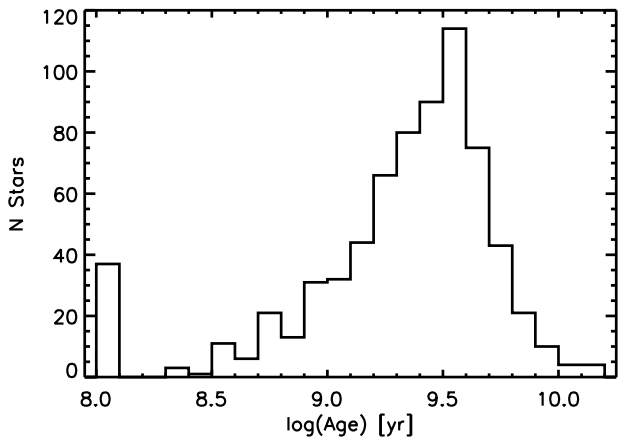}
\caption{ The distribution of ages found in our sample using the mass-age relation. }
\label{massagedist}
\end{figure}

\subsection{Mass-Age Relation}
\label{massagerelation}

We first derive ages from a simple giant branch mass-age relation. The PARSEC isochrones were used to find a relation between log(mass) and log(age) for all evolved stars. It should be noted that we work in $\log_{10}$(age), denoted as $\tau$, because the isochrones are given in a $\log_{10}$(age) grid. As the observed sample was selected for giants using a simple dereddened color cut, contributions from more advanced stages of evolution (e.g. RC, asymptotic giant branch (AGB)) are likely. As found in Section \ref{aspcap}, 324 stars were identified as RC stars. The ages of such stars can also be constrained by their mass as they are in the final stages of evolution. Therefore the mass-age relation was determined from all isochrone points with $\log g < 3.8$ instead of selecting only the giant branch points from the PARSEC `stage' parameter. The relation is also metallicity dependent, therefore the mass-age relation was fit for single metallicity populations with $-2.2<$[Fe/H]$<0.6$ in steps of 0.1~dex. The mass-age relation for a single metallicity was derived by applying a quadratic fit to the band of isochrone points with the given metallicity. The derived relations are shown in Figure \ref{massagefit} for a sample of metallicities as indicated. The isochrones are given in log(age) from 9.0 to 10.1 in steps of 0.05~dex. 

Given the spectroscopic metallicity of each star, the corresponding mass-age relation was applied to determine the age from the mass. As the isochrones cover an age range of 100~Myr to 12.6~Gyr, we assign an age of 100~Myr to stars with a mass greater than 5 $M_{\odot}$. We also assign an age of 14~Gyr to stars less massive than 0.8 $M_{\odot}$, limited by the age of the Universe. The distribution of ages determined through the simple mass-age relation is shown in Figure \ref{massagedist}.

The uncertainty in the age estimate is directly dependent on the error in the mass calculation. Using the derived mass-age relation, this results in a 0.38~dex uncertainty in log(age) for a $\log g$ uncertainty of 0.11~dex, which is a factor of two in age. The APOGEE errors in [M/H] are smaller than the isochrone [Fe/H] grid spacing and the dependence of the mass-age relation on [Fe/H] is weak compared to the isochrone grid spacing, therefore the [M/H] errors have a negligible effect on the age uncertainty.

Although the spread in the mass-age relation for all evolved stars is mainly due to the metallicity, there is also a small spread in the possible age for a single mass and metallicity. This spread is due to the lifetimes of each evolutionary stage and the contamination from non-RGB stars. The surface gravity, luminosity, and temperature of a single mass red giant star change as it ascends the giant branch. In addition, these parameters change as the star evolves through more advanced stages of evolution. The giant branch lifetime is between a few to 10\% of the total stellar lifetime, therefore some age precision can be gained if the star's exact position along the giant branch is known. This also allows the ages of stars doing more advanced stages of evolution, such as RC stars, to be more precisely determined based on their observed parameters. To get better age resolution, we test probabilistic isochrone matching techniques that use the individual measured parameters instead of the single combined parameter of mass, see sections \ref{prob} and \ref{hierarchical}.

\begin{figure}[t!]
\centering
\includegraphics[angle=90,width=0.45 \textwidth]{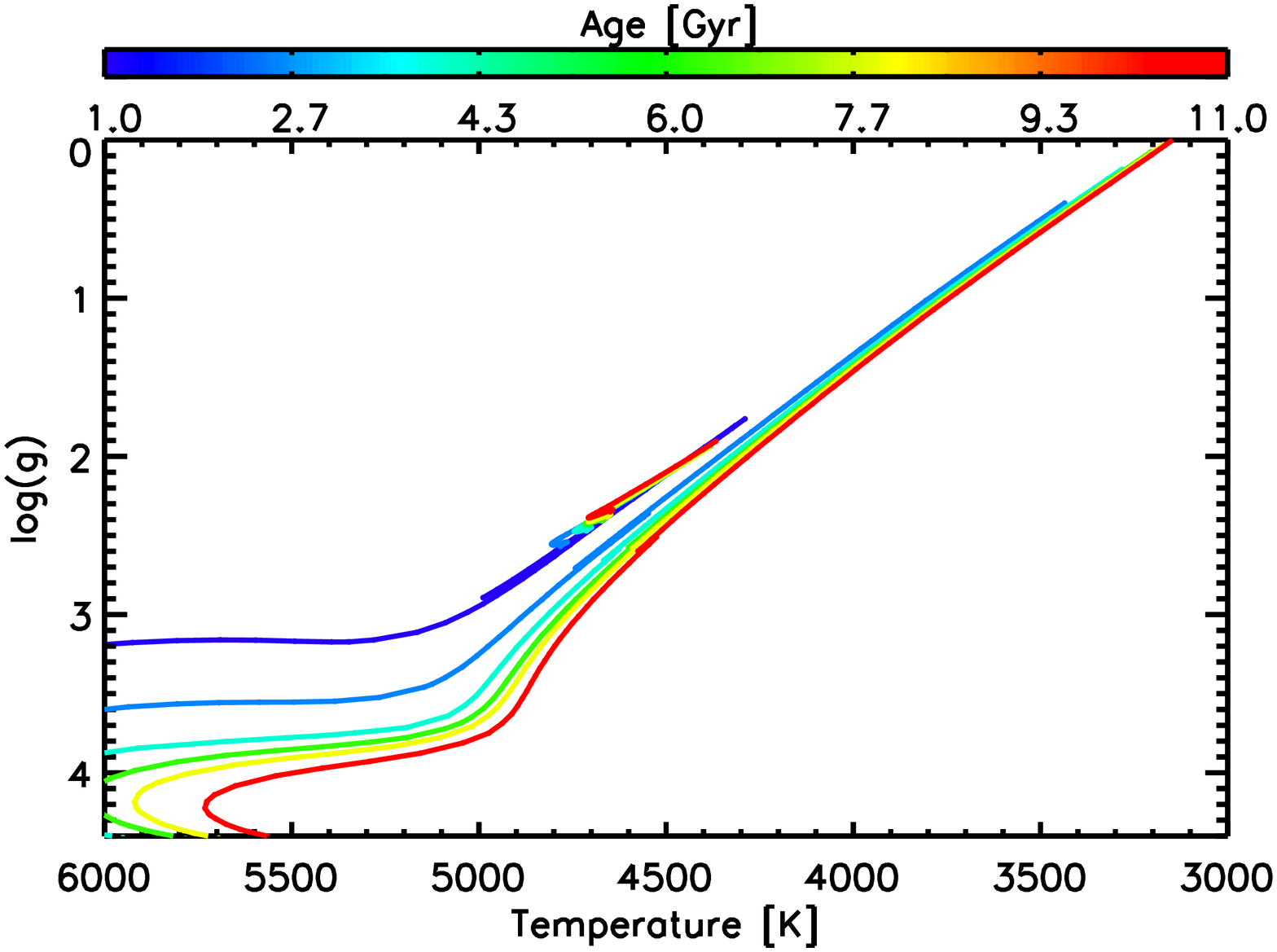}
\includegraphics[angle=90,width=0.45 \textwidth]{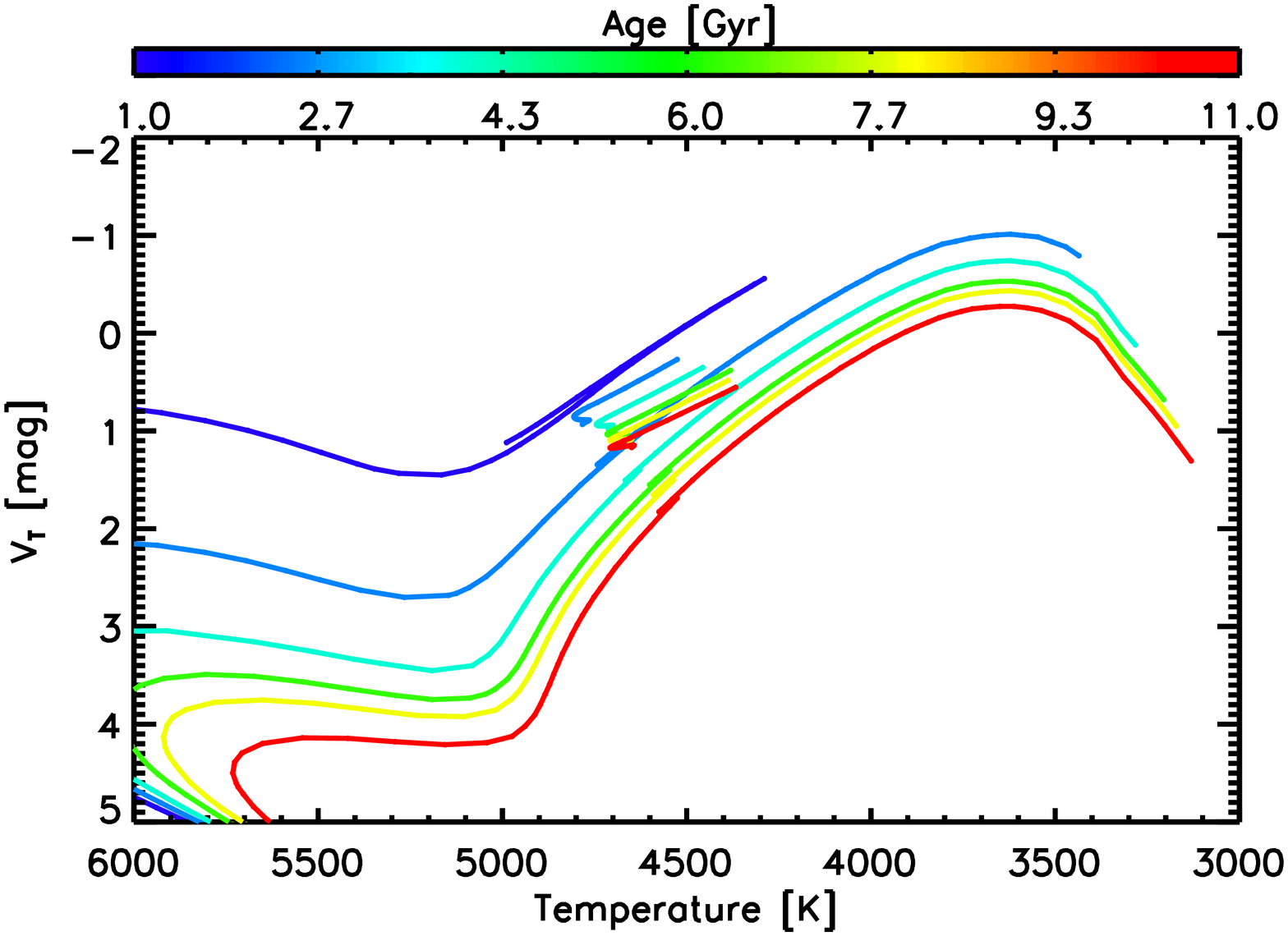}
\caption{ HR diagram of solar metallicity PARSEC isochrones in $\log g$ space (top) and $M_{V_T}$ magnitude space (bottom). The color represents age. There is a larger separation of the giant branch {and RC} with age in $M_{V_T}$ magnitude space than in $\log g$ space.}
\label{isoHR}
\end{figure}

\subsection{Isochrone Matching Age}
\label{prob}

If the atmospheric parameters are measured precisely, isochrone matching can be used to get some estimate of the age for each star, although not as precise as for subgiants. As discussed in Section \ref{mass}, if the mass of a giant branch star is known then the age can be constrained. However, in this section we use the measured parameters needed to derive a mass and compare them directly to isochrones. We can apply constraints to more properties of the models if the luminosity, temperature, and surface gravity are explicitly used to find the best matching isochrone points rather than folded into the mass value. Typically, isochrone matching uses color or temperature, and luminosity to compare to observations. The precise surface gravity measured from the high resolution APOGEE spectra allows the mass to indirectly constrain the isochrone matching. As can be seen in Figure \ref{isoHR}, giant branch and RC isochrones of different ages have a larger separation in absolute magnitude space than in surface gravity space. While surface gravity measurements provide the mass constraint, a measured distance is important in determining ages of red giant stars. For this sample we use both accurate absolute magnitudes from the Hipparcos data and the surface gravity from APOGEE high resolution spectra to match to isochrones. 

To find an age through isochrone matching, we evaluate the age probability density function (PDF) calculated using Bayesian estimation. Following the method described by \citet{Jorgensen2005} the age PDF is given by 
\begin{equation}
f(\tau,\zeta, m) \propto f_0(\tau,\zeta, m) L(\tau,\zeta, m) \nonumber
\end{equation}
where $f_0$ is the prior density function and $L$ is the likelihood function. Here, both functions depend on age, $\tau$, metallicity, $\zeta$, and mass, $m$.
Integrating over mass gives 
\begin{equation}
f(\tau,\zeta) \propto \int f_0(\tau,\zeta, m) L(\tau,\zeta, m) S(\tau,\zeta,m) \diff m
\label{PDFeq2}
\end{equation}
where $S(\tau,\zeta,m)$ is the selection function we use to account for the fact that our sample contains only evolved stars. For giant stars the mass and age are non-separable, as seen from the mass-age relation in Section \ref{massagerelation}. We define the selection function as $S(\tau,\zeta,m) = $ 1 for $\log g <  3.8$ and $(J - K)_0 > 0.5,$ and 0 elsewhere. The prior density function, $f_0$, depends on the SFH ($\psi(\tau)$), the initial mass function (IMF, $\xi(m)$), and the metallicity distribution function (MDF, $\phi(\zeta)$). In this work we have the metallicity of the star measured to within 0.05~dex, which is small compared to the width of the MDF in the solar neighborhood, therefore we simply assume a flat MDF across the measured metallicity uncertainty. We also assume that the IMF is independent of age and metallicity. We can therefore write Equation (\ref{PDFeq2}) as 
\begin{equation}
f(\tau,\zeta) \propto \psi(\tau) \phi(\zeta) \int L(\tau,\zeta, m) S(\tau,\zeta,m) \xi(m) \diff m . \nonumber
\end{equation}

The likelihood function, $L(\tau,\zeta,m)$, is the likelihood that a given isochrone point matches the observed star based on a set of measured parameters. The likelihood function is calculated as the product of each observable parameter compared to an isochrone point assuming Gaussian uncertainties. To build the likelihood PDF, $L(\tau,\zeta,m)$ is summed over all isochrone points. Integrated over the mass this has the form
\begin{eqnarray}
\label{likelihood}
L(\tau, \zeta) &\propto& \sum_{i=1}^{n} \exp \left(\frac{- ( X_{\mbox{\footnotesize 1,obs}} - X_{\mbox{\footnotesize 1,iso},i} )^2 }{ 2 \sigma^2_{\mbox{\footnotesize X{\tiny 1},obs}}} \right) \nonumber \\
&\times& \exp \left(\frac{- ( X_{\mbox{\footnotesize 2,obs}} - X_{\mbox{\footnotesize 2,iso},i} )^2 }{ 2 \sigma^2_{\mbox{\footnotesize X{\tiny 2},obs}}} \right) \\
&\times& \exp \left( ... \right) \, S(\tau_i,\zeta_i,m_i) \, \xi(m_i) \, \Delta m \nonumber
\end{eqnarray}
where $X$ is the measured stellar parameter, $\sigma_{\mbox{\footnotesize X,obs}}$ is the uncertainty in $X$, and $n$ is the number of isochrone points. In practice we only sum over isochrone points within $3 \sigma$ of all of the measured parameter value. We note that because the isochrones are given in 0.05 steps in $\log_{10}$(age), we also work in $\log_{10}$(age). For all equations given, $\tau$ denotes $\log_{10}$(age). We assume a Chabrier lognormal IMF \citep{Chabrier2001} as is provided within the PARSEC isochrones. 

The measured parameters considered were temperature ($T_{\mbox{\small eff}}$), metallicity ([M/H]), surface gravity ($\log g$), and absolute Tycho $V$-band magnitude ($M_{V_T}$). Metallicity is included in all cases. The PARSEC isochrones have solar \ad abundances, therefore we derive an adjusted metallicity for each star in the observed Hipparcos sample to account for \ad abundance. It has been shown that \ad enhanced stars appear cooler than stars with solar \ad abundance of the same age and [Fe/H] \citep[see e.g. ][]{Salaris1993}. If ignored, this effect would result in an older age assigned to an \ad enhanced star. As this is exactly the trend suggested by chemical evolution models of the thick disk populations, it is crucial to account for any \ad enhancements when comparing to solar abundance isochrones. To correct for the temperature shift due to the stellar \ad enhancement, an adjusted [M/H] was used to compare to the isochrone [Fe/H]. We use the correction described in \citet{Salaris1993}. As our sample does not contain many \ad enhanced stars, only a small adjustment is required. For 94 \% of the observed sample the adjustment to the metallicity is less than 0.1~dex, the PARSEC isochrone [Fe/H] step size. The effect on the age estimate is almost always less than 0.1~dex. We discuss below the combination of measured parameters that allows for the most accurate age determination of red giants. For the purpose of comparing to a direct age estimate from the mass (see sections \ref{mass} and \ref{massagerelation}) we also consider the ages derived with this method using [M/H] and mass as the measured parameters.

Using $L(\tau,\zeta)$ as given in Equation (\ref{likelihood}), we assume a flat MDF over the uncertainty in [M/H] and integrate over metallicity to obtain the full age PDF for a single star.
\begin{equation}
f(\tau) \propto \psi(\tau) L(\tau). \nonumber
\end{equation}
The use of a grid of isochrones spaced in log(age) imposes a default prior of a flat SFH prior in log(age) or $\tau$. We adopt a flat SFH prior in age by weighting each $\tau$ bin of the PDF by the linear age of that bin. This results in the full age PDF of a single star. In Section \ref{hierarchical} we model the SFH as a parameterized function.

\subsection{Mock Data Tests}
\label{simtests}

Using a sample of simulated stars, we examine which combination of measured parameters results in the most accurate age determination. The simulated sample was created by selecting random points from the PARSEC isochrones to create a sample with a flat distribution in age. The sample contains 1200 stars with parameters $3500 \, \mbox{K} < T_{\mbox{\small eff}} < 5400 \, \mbox{K} $, $-1.0 < [\mbox{Fe/H}] < 0.7$, $3.8 < \log g < 0$, and $(J-K)_0 > 0.5$. This contains only evolved stars, covers a large range of metallicities, and includes the color cut imposed on the observational sample. Gaussian noise was introduced to the points based on the typical observational uncertainties of our sample, 92~K in $T_{\mbox{\small eff}}$, 0.11~dex in $\log g$, 0.05~dex in [M/H], and 0.15~mag in $M_{V_T}$. From these parameters a spectroscopic mass was calculated just as is done with the observed sample. The disagreement in the true mass of the isochrones point and the mass calculated from the parameters with added noise is 38 \%, in agreement with the estimate based on observational errors in Section \ref{mass}.

\begin{figure}[t]
\centering
\includegraphics[angle=90,width=0.23 \textwidth]{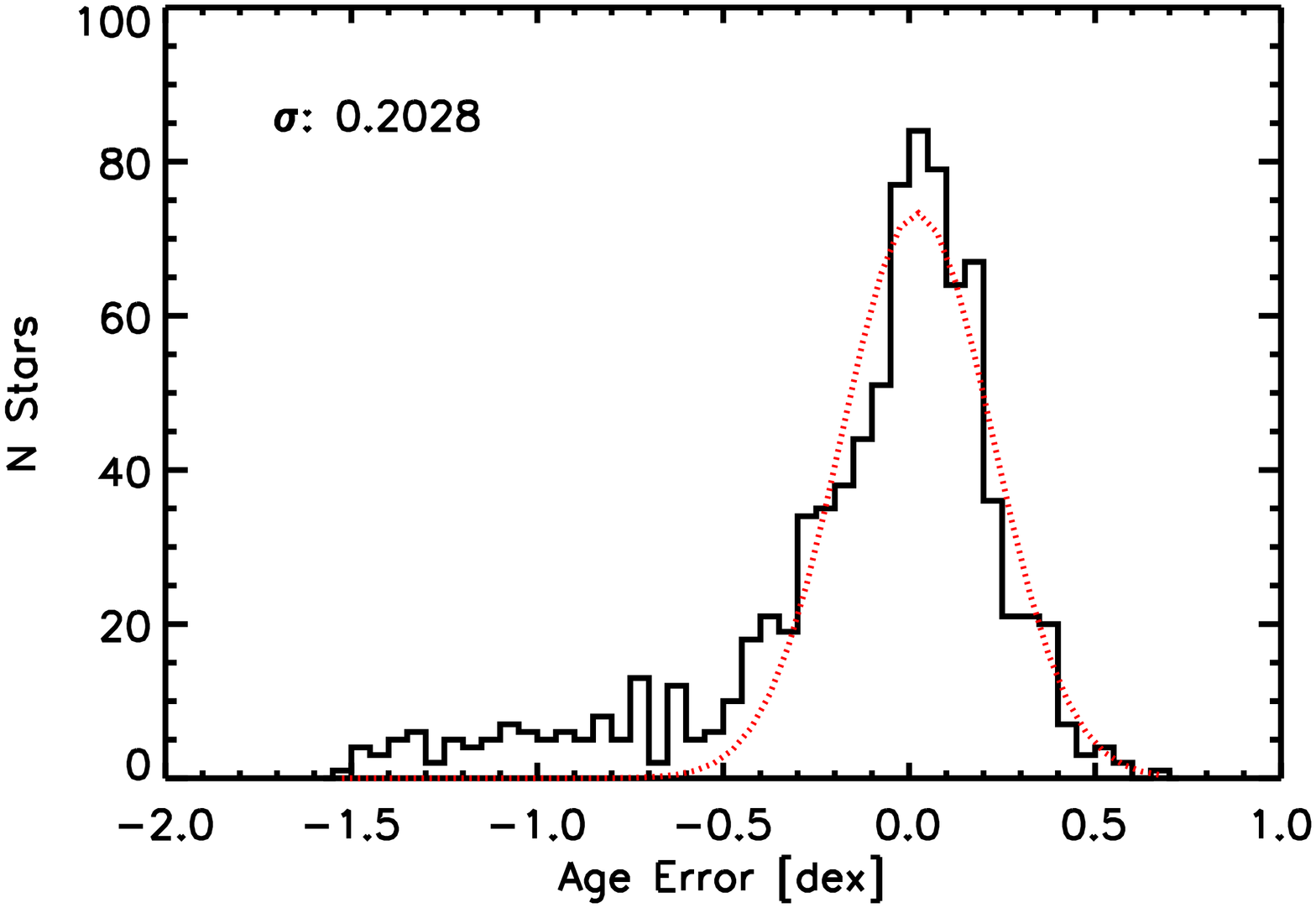}
\includegraphics[angle=90,width=0.23 \textwidth]{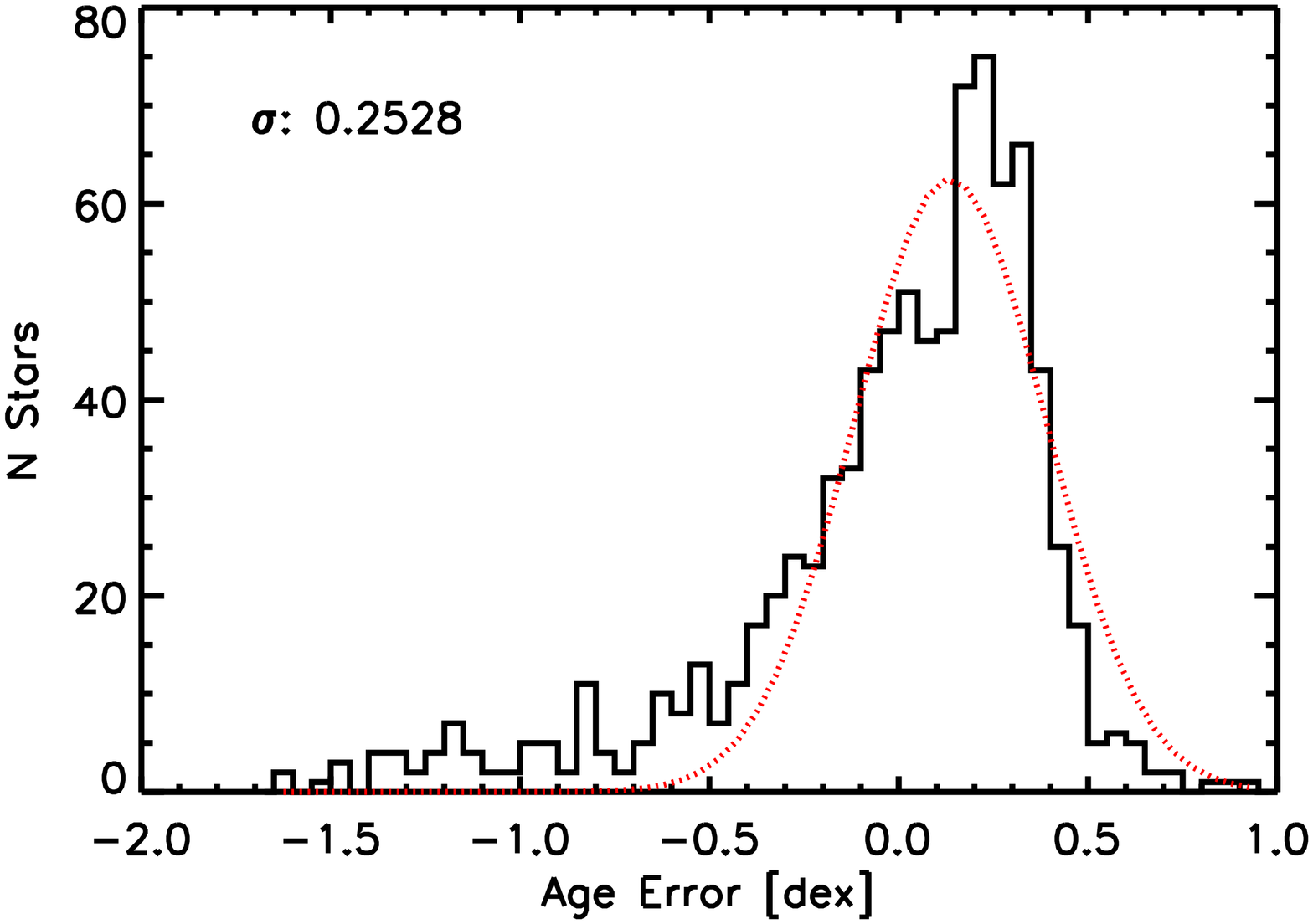}
\includegraphics[angle=90,width=0.23 \textwidth]{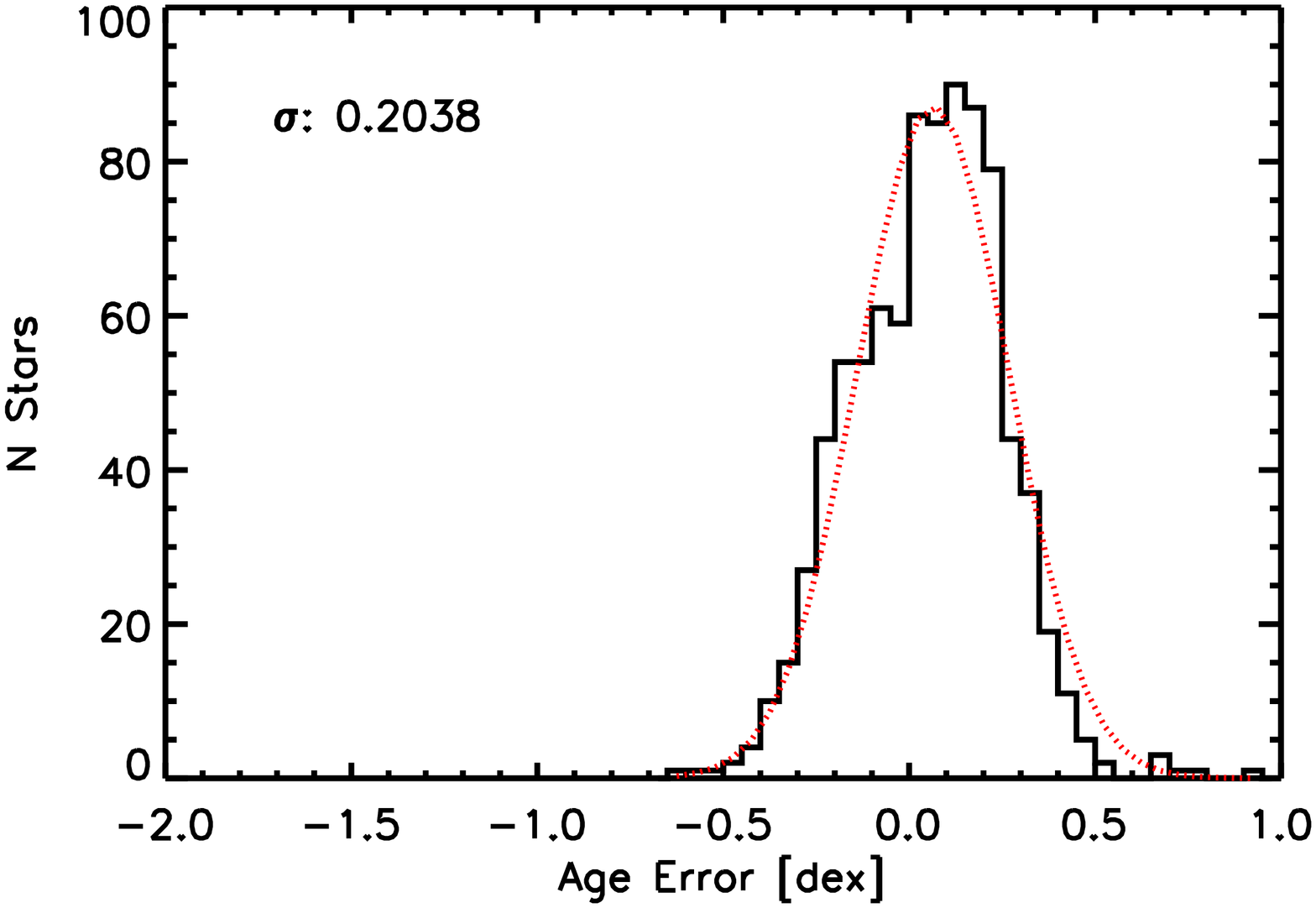}
\includegraphics[angle=90,width=0.23 \textwidth]{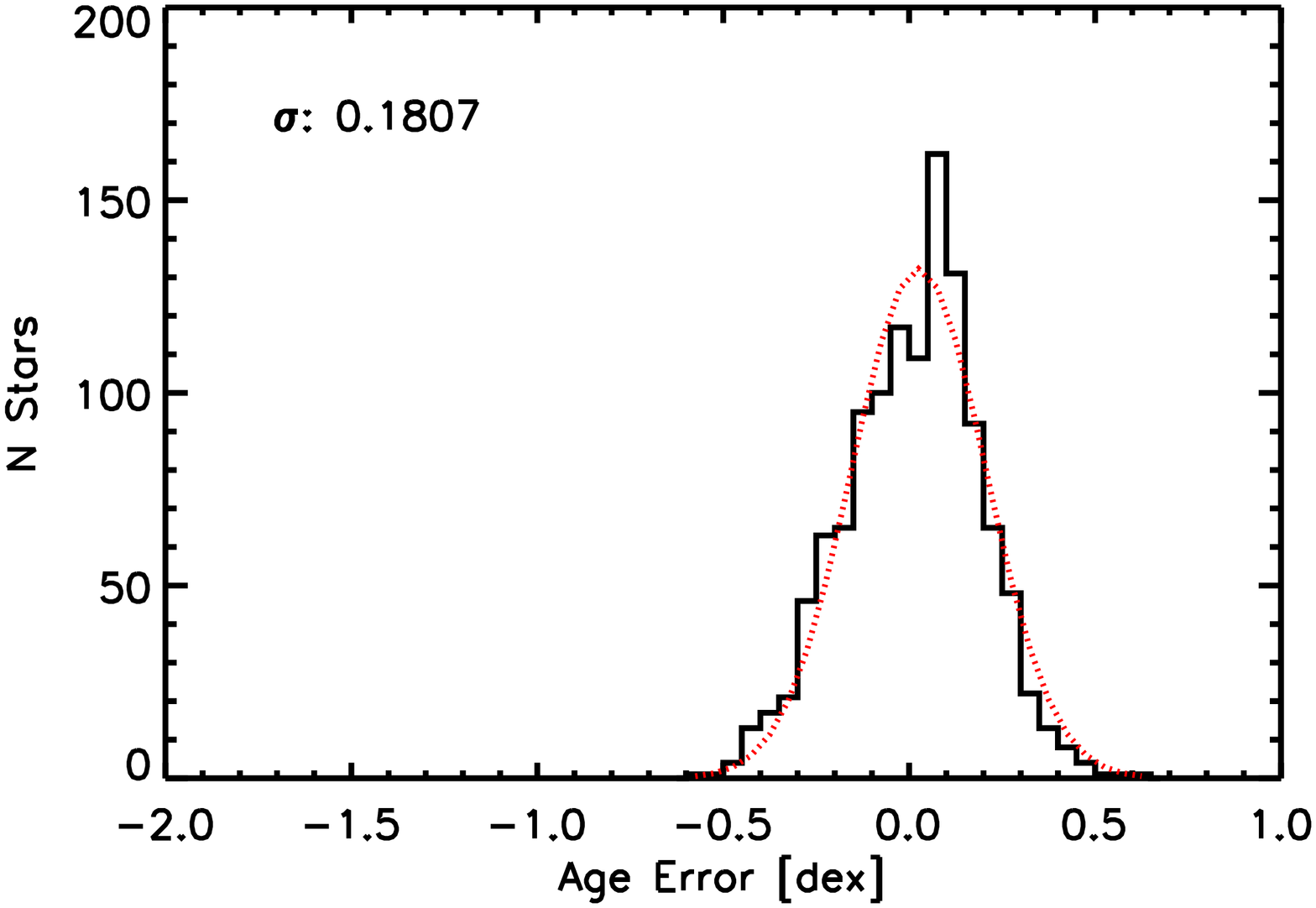}
\caption{ The distribution of the age error (real minus recovered) using a variety of measured parameters. 
[Fe/H] and $M$ (top left), 
[Fe/H], $T_{\mbox{\small eff}}$, and $\log g$ (top right),   %
[Fe/H], $T_{\mbox{\small eff}}$, and $M_{V_T}$ (bottom left),  
[Fe/H], $T_{\mbox{\small eff}}$, $M_{V_T}$, and $\log g$ (bottom right). 
The recovered age is more accurate when $M_{V_T}$ is included in the isochrone match. The top panels both show a tail of recovered ages that are much larger than the true age. The age error distribution is similar for the bottom panels, although the bottom right panel has the smallest standard deviation.} 
\label{ageerror}
\end{figure}

\begin{figure}[t]
\centering
\includegraphics[angle=90,width=0.23 \textwidth]{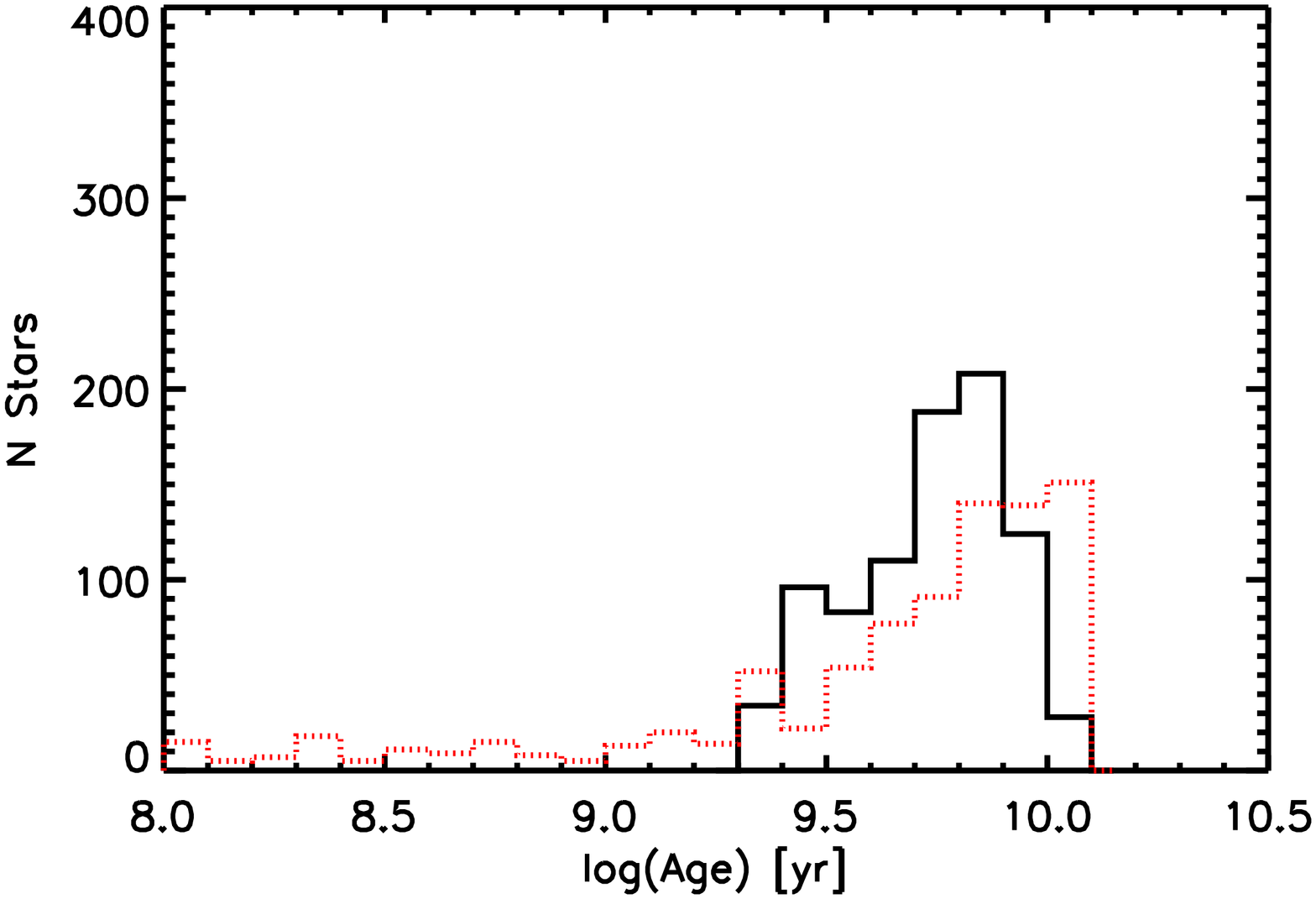}
\includegraphics[angle=90,width=0.23 \textwidth]{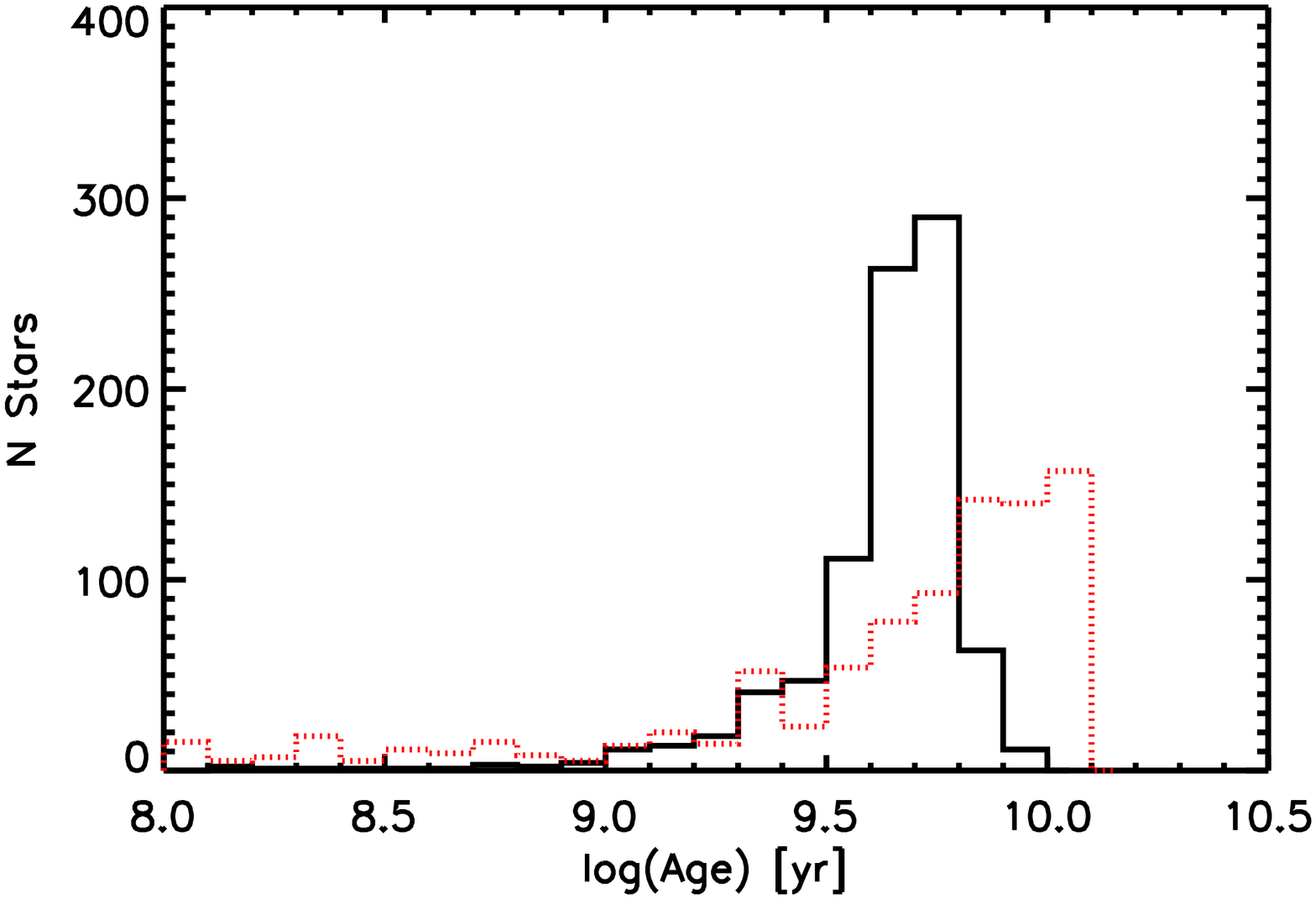}
\includegraphics[angle=90,width=0.23 \textwidth]{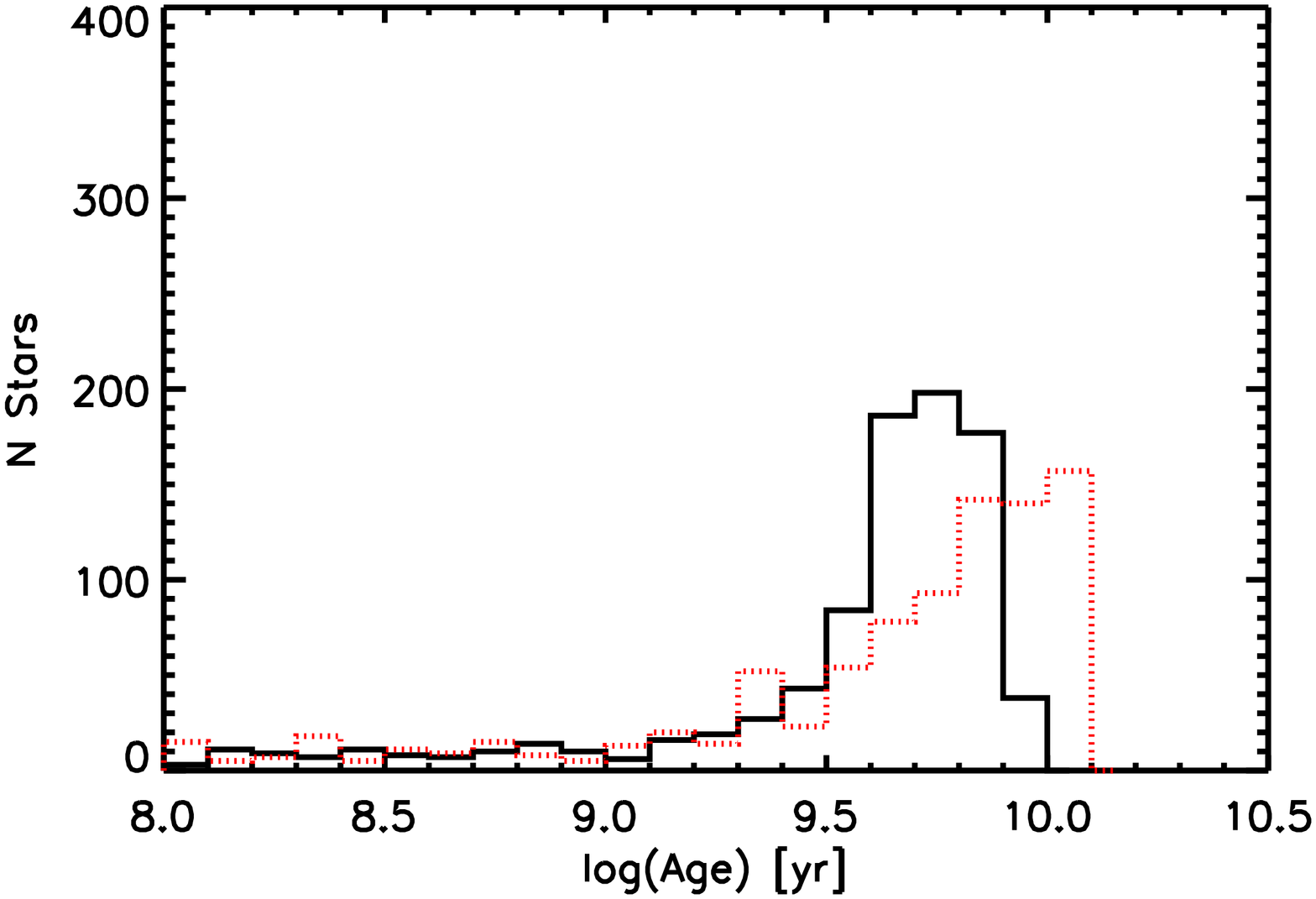}
\includegraphics[angle=90,width=0.23 \textwidth]{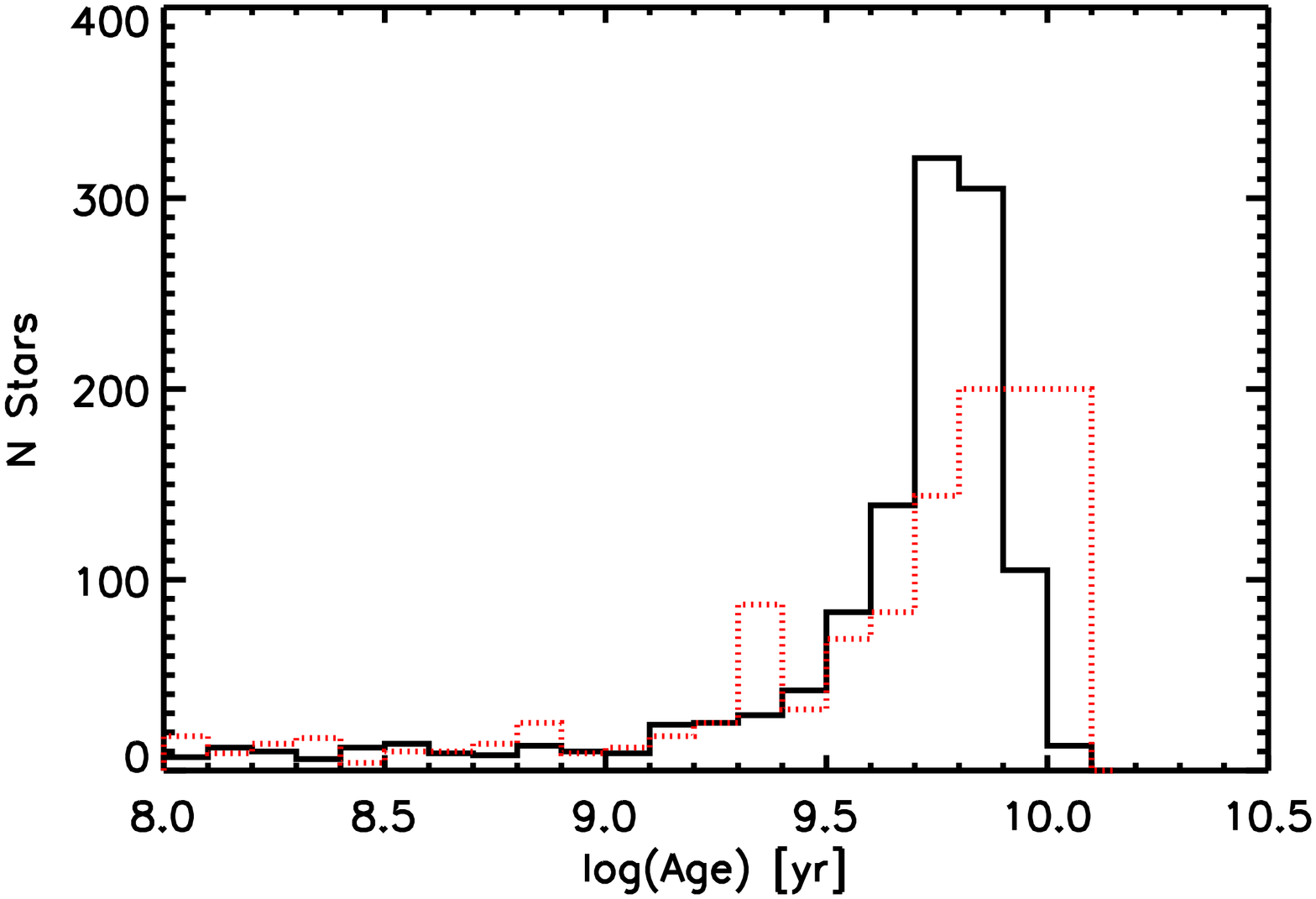}
\caption{ The distribution of the recovered ages from the simulated sample found using 
[Fe/H] and $M$ (top left), 
[Fe/H], $T_{\mbox{\small eff}}$, and $\log g$ (top right), 
[Fe/H], $T_{\mbox{\small eff}}$, and $M_{V_T}$ (bottom left), 
[Fe/H], $T_{\mbox{\small eff}}$, $M_{V_T}$, and $\log g$ (bottom right).  
The red dashed line shows the distribution of true ages. The top panels are highly peaked around 4-6~Gyr. The bottom panels recover younger ages for the older stars, despite the narrower error distribution.
}
\label{agedist}
\end{figure}

\begin{figure}[t]
\centering
\includegraphics[angle=90,width=0.45 \textwidth]{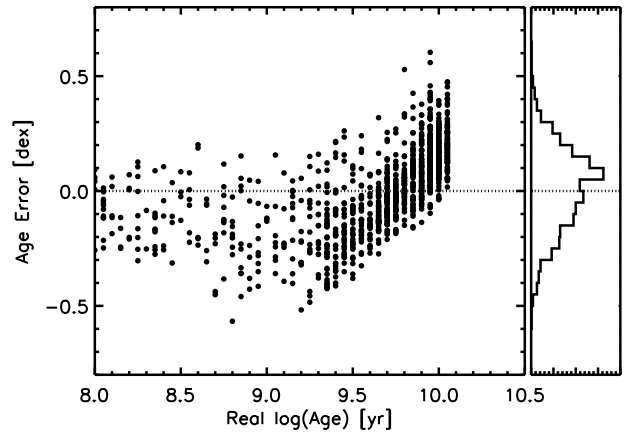}
\caption{ Age error as a function of true age for the simulated sample found using [Fe/H], $T_{\mbox{\small eff}}$, $M_{V_T}$, and $\log g$ as measured parameters for a Bayesian analysis. Using this method of age estimation, the young stars are assigned ages that are preferentially too old and the old stars are assigned ages that are preferentially too young. The right panel shows the histogram of age errors as shown in the lower right panel of Figure \ref{ageerror}.}
\label{ageverror}
\end{figure}

An age PDF was calculated for each star in this simulated sample using the method described in Section \ref{prob} for several combinations of measured parameters, however, [Fe/H] was included in all cases. To assign a single age to a star we take the mean of the age PDF. The age PDF for an evolved star is often not a Gaussian PDF, and can have more than one local maximum because these stars commonly have measured parameters similar to older or younger stars at a different stage of evolution (e.g. RC or AGB stars). The mean of the PDF is more sensitive to multiple peaks, which can introduce larger uncertainties in the age of a single star, but can result in a more accurate age distribution for a large sample. The difference in log(age) between the real and recovered age values for a few test parameter sets is shown in Figure \ref{ageerror}. In these figures, a Gaussian is fit to the distribution of errors to determine the $1 \sigma$ errors. Both the Gaussian fit and the $\sigma$ value are indicated in the figure.

The upper left panel of Figure \ref{ageerror} shows the age errors of a Bayesian isochrone matching analysis using [Fe/H] and $M$ as the measured parameters. This analysis compares directly with the mass-age relation analysis from Section \ref{massagerelation}. The age errors here are generally smaller than the uncertainties from the mass-age relation due to the inclusion of priors, however, there is a tail of large age errors. If we compare this analysis with an analysis based directly on the individual parameters used to calculate the mass ([Fe/H], $T_{\mbox{\small eff}}$, $M_{V_T}$, and $\log g$, lower right), the error in recovered ages is larger for the analysis based on the mass-age relation alone (upper left). This is due to the large uncertainty in the mass with no explicit constraints on other measured parameters. Analyses based on HR diagram position ([Fe/H], $T_{\mbox{\small eff}}$, and $M_{V_T}$, lower left) and [Fe/H], $T_{\mbox{\small eff}}$, and $\log g$ (upper right) illustrate the importance of including a constraint on luminosity. Due to the larger separation of RGB isochrones in $M_{V_T}$ than in $\log g$, see Figure \ref{isoHR}, the distribution of age errors is always narrower when $M_{V_T}$ is included in the isochrone matching. The top panels contain no direct magnitude comparison and show large tails of recovered ages that are too old, indicating that the absolute magnitude gives much better age resolution in giants, particularly at young ages. 

For the cases in which absolute magnitude is included in the isochrone matching and the full error distribution is smallest, the largest individual age errors occur in stars with $\log g \sim 2.4$. This region of the HR diagram contains not only RGB stars, but also RC and AGB stars. \citet{Anders2015} note that in their analysis, this region contains the most stars with double-peaked PDFs. An isochrone probability match based on all observed parameters ([Fe/H], $T_{\mbox{\small eff}}$, $M_{V_T}$, and $\log g$) was found to give the smallest errors with a distribution width of 0.18 dex. 

\begin{figure}[t]
\centering
\includegraphics[angle=90,width=0.35 \textwidth]{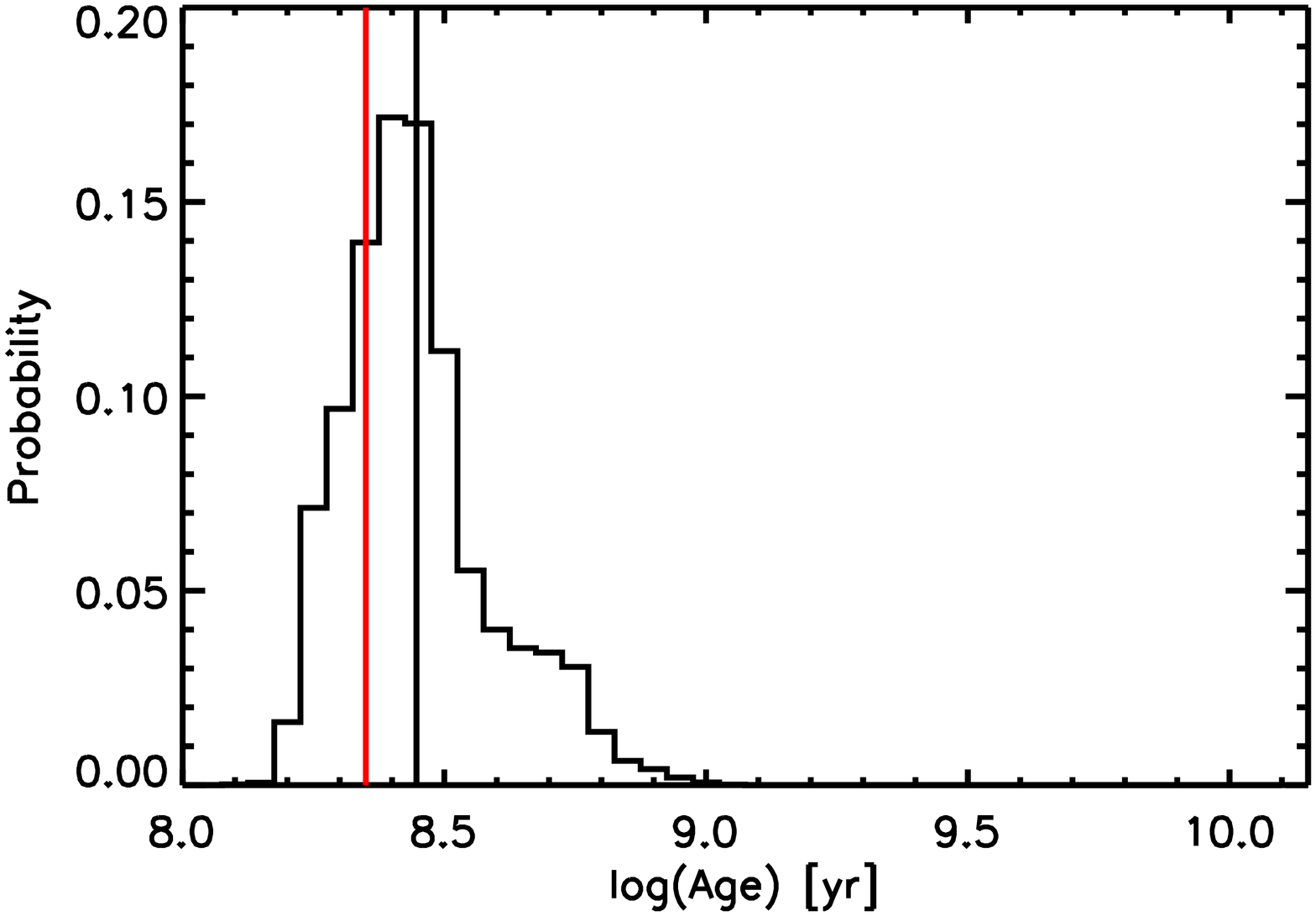}
\includegraphics[angle=90,width=0.35 \textwidth]{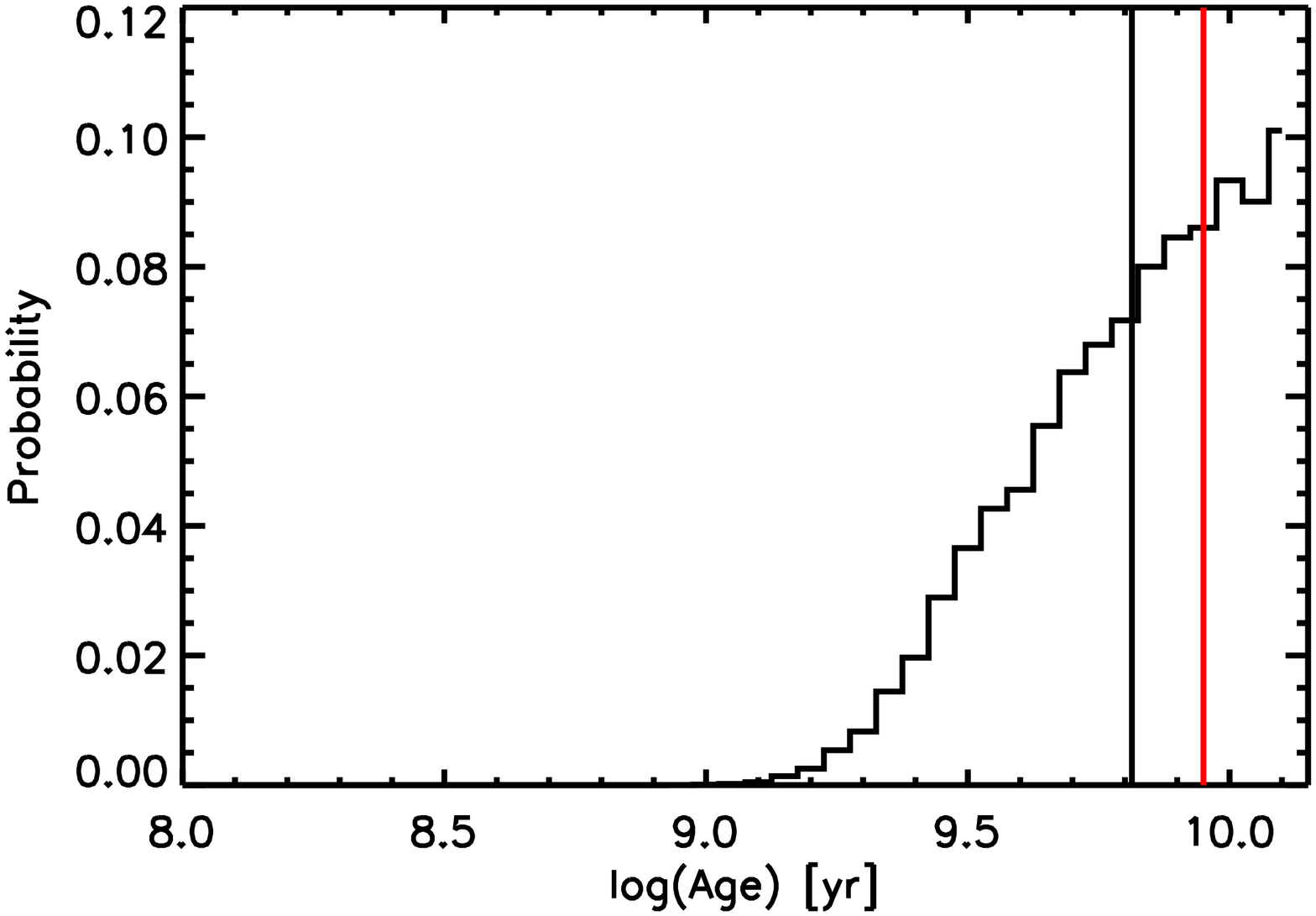}
\caption{ Example age PDFs of a star with $\tau = $ 8.35 (top) and 9.95 (bottom). The PDF of the older star is truncated at 10.1 due to encountering the isochrone grid edge. The true age is indicated by the red vertical line and the mean of the PDF is indicated by the black vertical line.}
\label{exPDF}
\end{figure}

The distribution of recovered ages for each test case is shown in Figure \ref{agedist} as the black solid line compared to the true age distribution of the sample as the dotted red line. It is again clear that when absolute magnitude is included in the isochrone matching, the youngest ages are recovered. However, the oldest ages are not recovered, even if $M_{V_T}$ is included. Figure \ref{ageverror} shows the age error as a function of true age. In this Figure it is apparent that the ages of young stars are over estimated, while the ages of old stars are significantly underestimated. We believe these biases are a consequence of taking the mean of the PDF to be the age of the star, as seen in Figure \ref{exPDF}. 

\begin{figure}[t]
\centering
\includegraphics[angle=90,width=0.4 \textwidth]{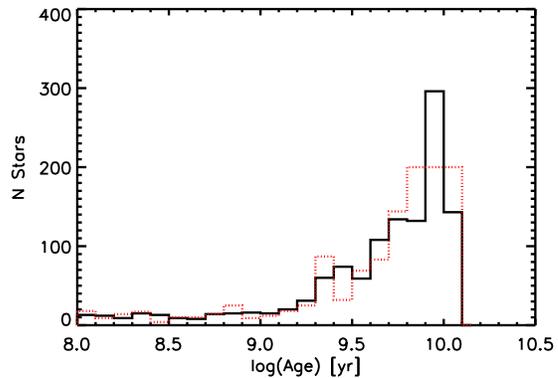}
\caption{The distribution of the recovered ages from the simulated sample found using [Fe/H], $T_{\mbox{\small eff}}$, $M_{V_T}$, and $\log g$ and uncertainties one order of magnitude smaller than the typical observational uncertainties shown as the solid black line. In comparison, the dotted red line indicates the true age distribution of the input sample. We find that if the introduced uncertainties are very small, the distribution of recovered ages matches the input distribution.}
\label{noerrdist}
\end{figure}

Figure \ref{exPDF} shows two examples of the age PDFs from the simulated sample. The top panel is a $\tau=8.35$ star and the bottom panel is a $\tau=9.95$ star. For young stars, the PDF often has a tail towards older ages. Taking the mean of this distribution can push the age of a very young star older. For the older ages, the bias towards younger ages is likely strongly driven by the isochrone grid edge. The PDF of the older star is peaked at old ages, but is truncated at the isochrone grid edge, pushing the mean of the PDF towards younger ages. In addition, $\Delta \tau$ is not linear with $\Delta$ observational uncertainty, and the separation between isochrones in $T_{\mbox{\small eff}}$, $M_{V_T}$, and $\log g$ decreases with increasing age. This results in larger uncertainties in age at older ages for the same uncertainty in $T_{\mbox{\small eff}}$, $M_{V_T}$, or $\log g$. We test analyzing a simulated sample using uncertainties that are an order of magnitude smaller than our typical observational uncertainties. Figure \ref{noerrdist} shows that when the imposed uncertainties are very small, the distribution of recovered ages matches the true age distribution of the input sample. From this we conclude that the bias in recovered age is a consequence of taking the mean on the PDF, the non-linear relation between $\tau$ and the measured parameters, and because the isochrone grid edge truncates the PDFs at $\tau$ of 10.1. This bias is most significant at the oldest age bin. 

\subsection{Hipparcos Sample}
\label{hipages}

\begin{figure}[t]
\centering
\includegraphics[angle=90,width=0.4 \textwidth]{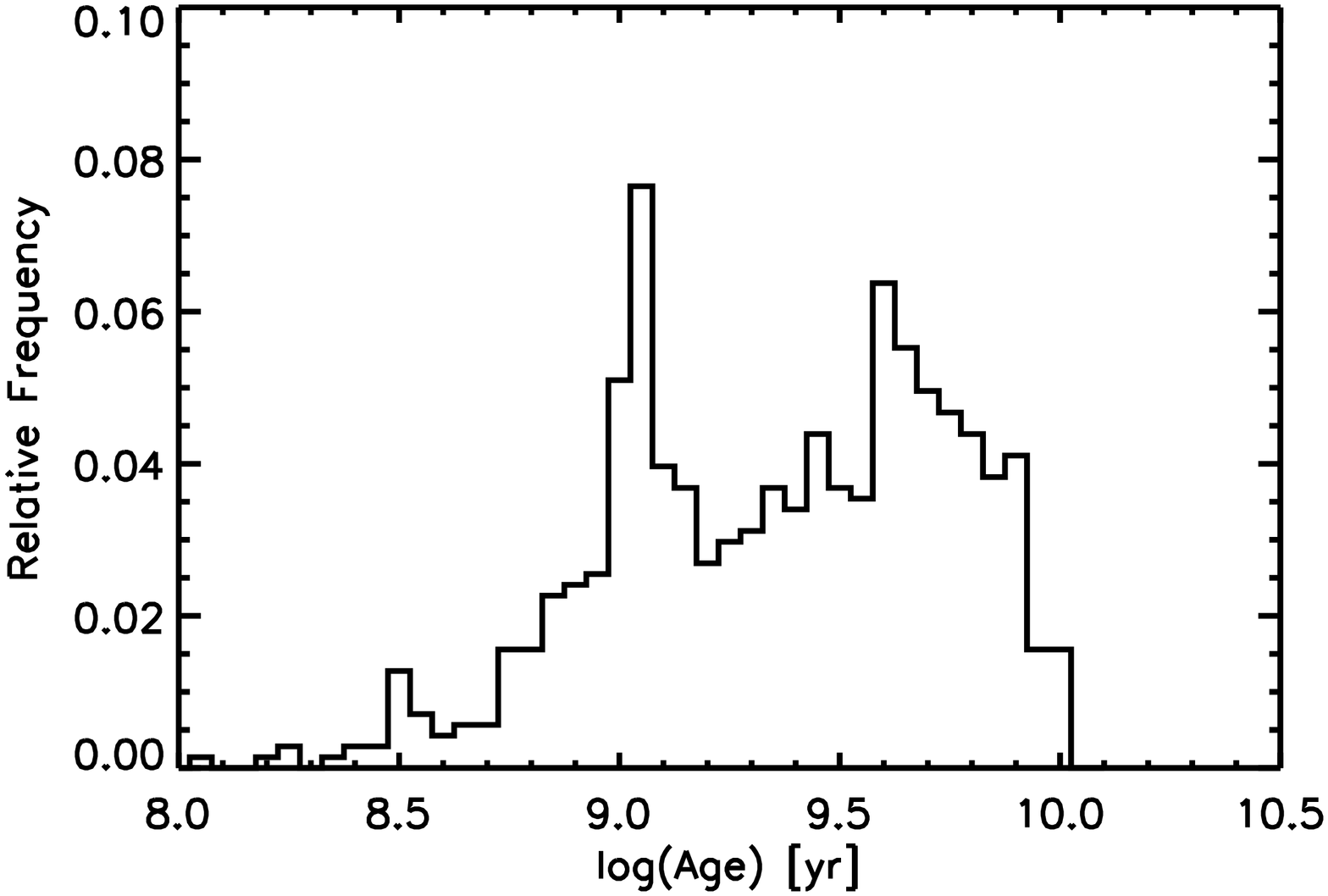}
\includegraphics[angle=90,width=0.4 \textwidth]{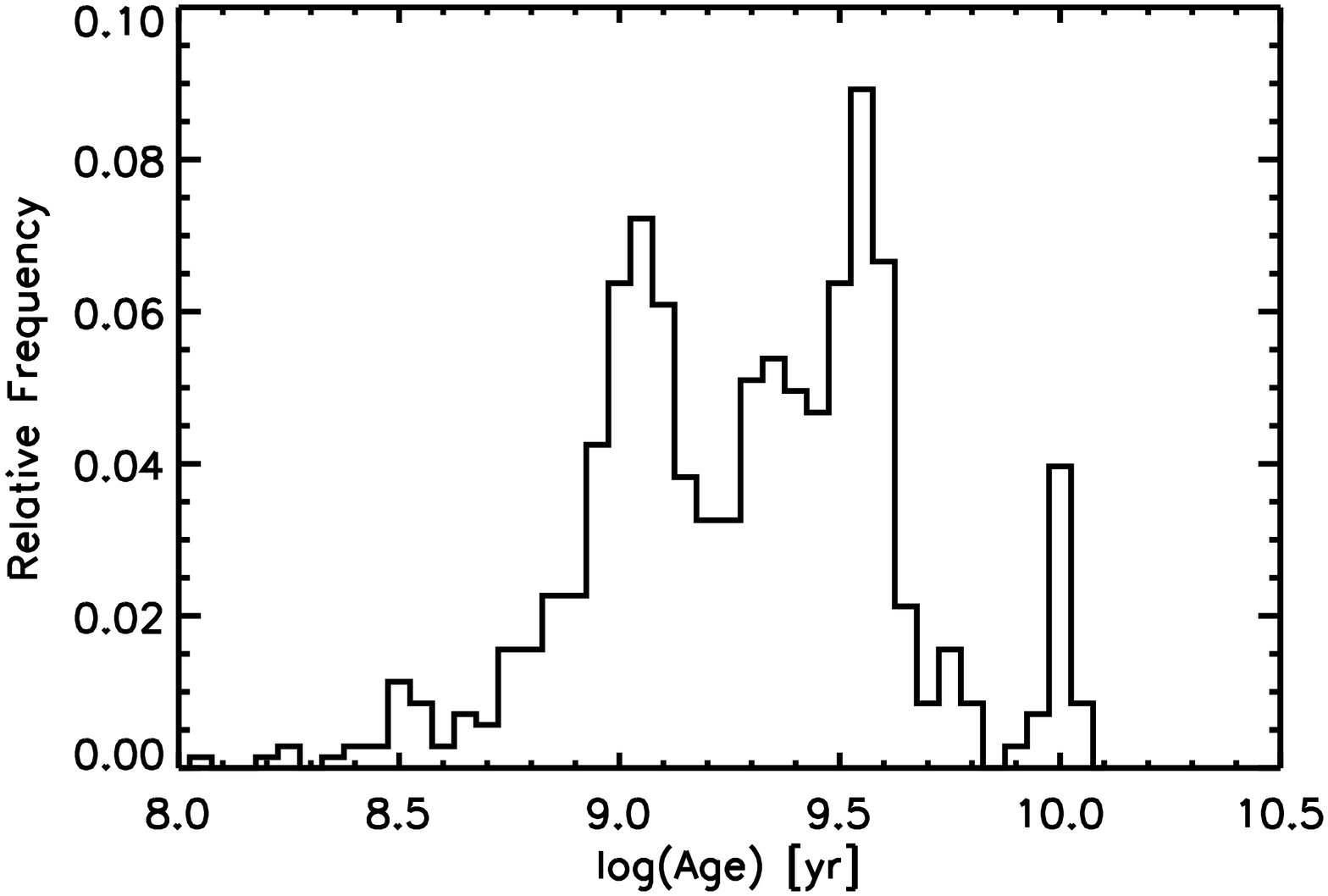}
\caption{ The distribution of ages determined for the APOGEE Hipparcos sample using a Bayesian approach (top) and the empirical Bayesian analysis using the model SFH as a prior (see Section \ref{hierarchical}, bottom). The age distribution is skewed towards younger ages with a young peak at 9.1 and an older peak at 9.6 for the Bayesian ages and 9.55 for the empirical Bayesian ages. }
\label{agedisthip}
\end{figure}

\begin{figure}[t]
\centering
\includegraphics[angle=90,width=0.42 \textwidth]{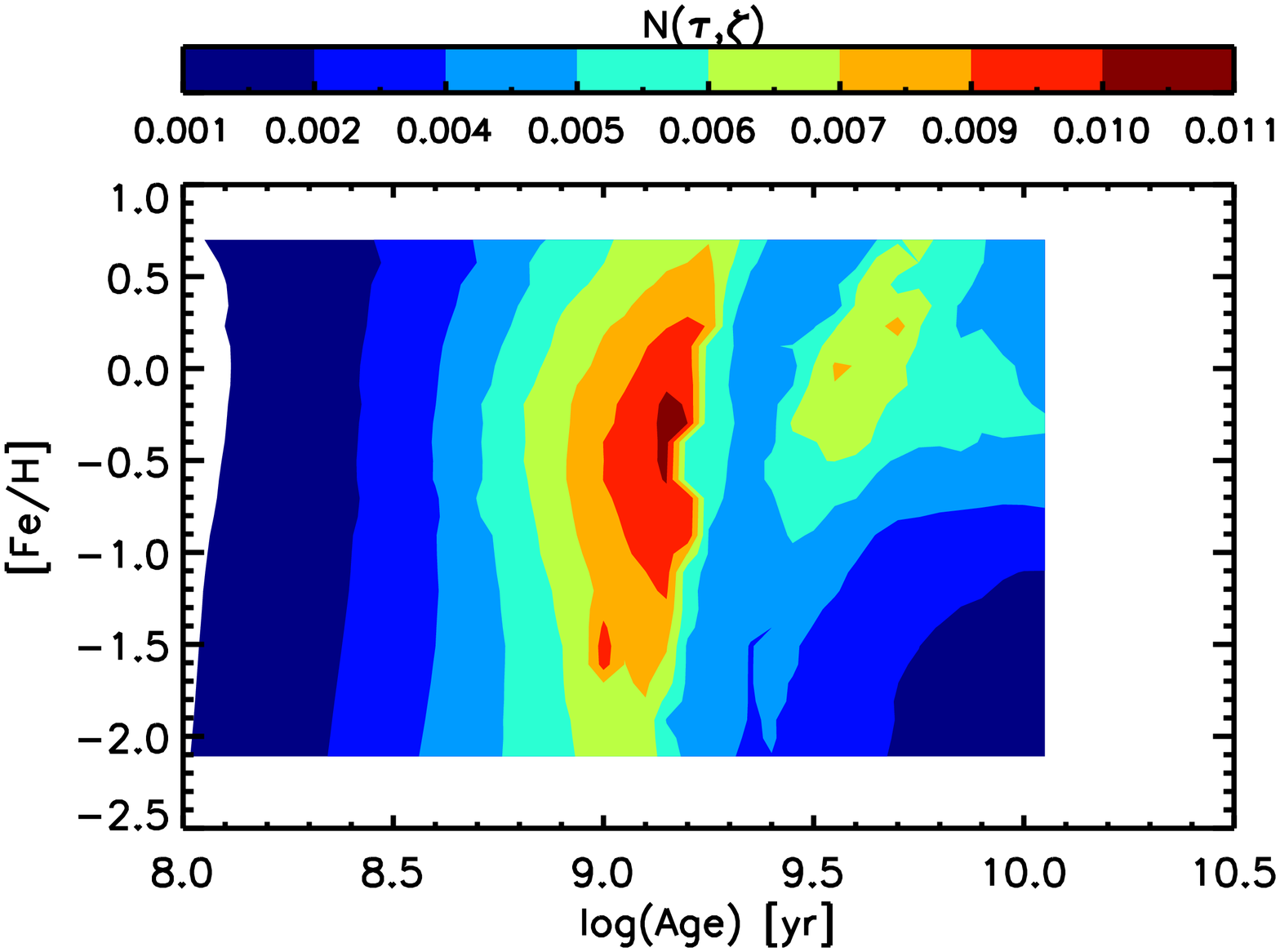}
\includegraphics[angle=90,width=0.39 \textwidth]{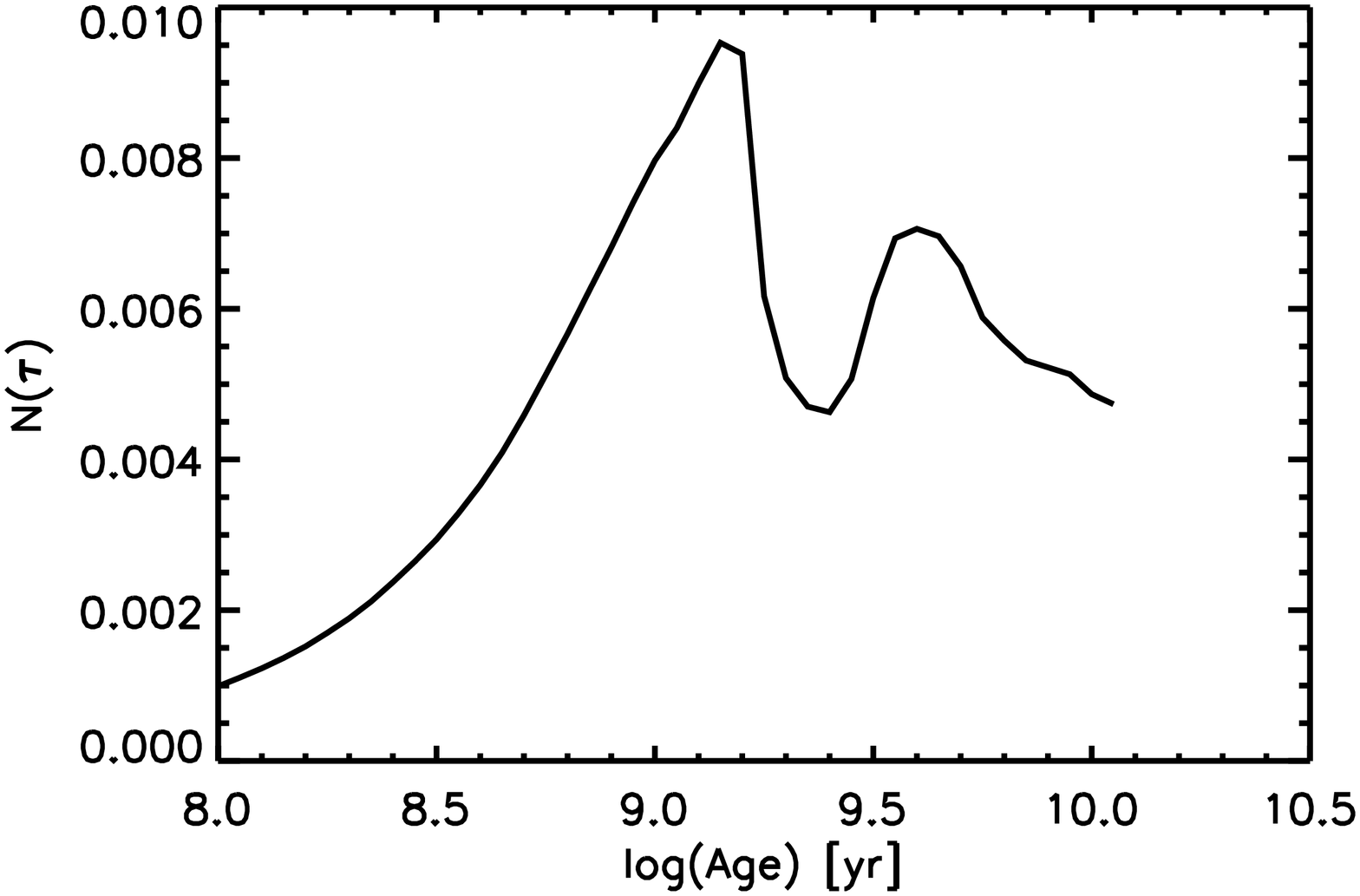}
\caption{ The expected distribution, $N(\tau,\zeta)$, for a sample of evolved stars selected with the same criteria as the local red giant sample in age, $\tau$, and metallicity, $\zeta$ (top). We also examine the expected age distribution, $N(\tau)$, assuming a solar neighborhood MDF similar to that found by \citet[][bottom]{Hayden2015}.}
\label{Ntau}
\end{figure}

Using the isochrone matching method described above, we determine individual ages for the sample of nearby giant stars observed with the 1m+APOGEE. The measured parameters used to compare to the isochrones were the \ad adjusted [M/H] described in Section \ref{prob}, $T_{\mbox{\small eff}}$, $M_{V_T}$, and $\log g$. The age distribution of the sample is shown in Figure \ref{agedisthip}. This distribution has a peak at $\tau = 9.1$ and at $\tau = 9.6$. Note that this would be expected even for a population of giants with a constant SFH due to the evolutionary rate of stars of different masses. The evolutionary rate of high mass stars is higher at younger ages and the distribution of RC lifetimes has an excess at $\sim 2$~Gyr ($\tau = 9.3$) with respect to the RGB. This results in an age distribution that is skewed towards younger ages. Figure \ref{Ntau} shows the expected age and metallicity distribution for a sample of evolved stars with our selection function, $S(\tau,\zeta,m)$. In this figure, $N(\tau,\zeta)$ is calculated from the isochrones as 
\begin{equation}
N(\tau,\zeta) = \int \xi(m) S(\tau,\zeta,m) \diff m . \nonumber
\end{equation}
We also calculate $N(\tau)$ assuming a MDF for the solar neighborhood similar to that found by \citet{Hayden2015}. There is a strong peak in $N(\tau)$ at $\tau$ of 9.2 from RC and AGB stars, and a smaller peak at $\tau$ of 9.6 from RGB stars. The age distribution of the observed sample is has a similar behavior to this expected age distribution based on our selection function and the isochrones. We do note that the age of the peaks in the observed sample are slightly offset from the theoretical prediction, complicating analysis of the underlying population. However, the peaks are within 0.1~dex of the theoretical prediction, which is a reasonable result given the uncertainty of 0.18~dex in the simulated sample.

We use the variance of the age PDFs as the uncertainty in the age for each star. The mean uncertainty in the age using this Bayesian method is 0.15~dex, with a spread of about 0.07~dex. This about half the uncertainty in age using the mass-age relation.

\subsection{Hierarchical modeling}
\label{hierarchical}

The derived ages described above in Section \ref{prob} assume a flat SFH in $\tau$ with no metallicity dependence. However, we do not know the true SFH of this sample. In this section we explore a Bayesian hierarchical modeling approach to determining a more realistic prior, $f_0(\tau,\zeta,m)$, for this sample. The hierarchical modeling approach uses the likelihood age PDFs from Equation (\ref{likelihood}) to determine the parameters of a model SFH. In this way we use the full age PDF of each star in the whole sample to inform our SFH rather than independently finding individual stellar ages assuming a flat SFH. Although one could adopt a SFH based on previous knowledge of the solar neighborhood, a hierarchical modeling approach allows the data to inform the SFH of the given sample. A Bayesian hierarchical analysis could be applied to a sample outside the solar neighborhood, where there are fewer known constraints on the SFH.

In this method, we model the prior, $f_0(\tau,\zeta,m | a)$, as a function with free parameter(s) $a$. The PDF for a given parameter $a$ is then
\begin{equation}
p(a | \mbox{data}) \propto p(\mbox{data} | a) \, p(a). \nonumber
\end{equation}
This can be re-written in terms of the model prior and the likelihood distributions from Section \ref{prob},
\begin{eqnarray}
&\propto& p(a) \int p(\mbox{data} | \tau, \zeta, m) \, p(\tau,\zeta,m | a) \diff \tau \diff \zeta \diff m \nonumber \\
&\propto& p(a) \int p(\mbox{data} | \tau, \zeta, m) \, N(a) \nonumber \\
&\times& f_0(\tau,\zeta,m | a) \diff \tau \diff \zeta \diff m \nonumber
\end{eqnarray} 
where $N(a)$ is a model normalization term defined such that 
\begin{equation}
N(a) \int f_0(\tau,\zeta,m) \, S(\tau,\zeta,m) \diff \tau \diff \zeta \diff m = 1. \nonumber
\end{equation}
This normalization term ensures that the total probability that a star in the sample is a giant given the model prior is unity. This is necessary as the sample is defined to contain only giants. Recall from Section \ref{prob} that the prior depends on the SFH, the IMF, and the metallicity distribution, as well as the selection function, $S(\tau,\zeta,m),$ to account for a sample of all giants. As we are modeling the SFH it now also depends on the parameter(s) $a$ of the model. Therefore, 
\begin{eqnarray}
p(a|\mbox{data}) &\propto& p(a) \int p(\mbox{data} | \tau, \zeta, m) \, \phi(\zeta) \, \xi(m) \nonumber \\
&\times& S(\tau,\zeta,m) \, N(a) \, \psi(\tau | a) \diff \tau \diff \zeta \diff m . \nonumber
\end{eqnarray}
Here $p(\mbox{data} | \tau, \zeta, m)$ is the likelihood function $L(\tau,\zeta,m)$ for an individual star, so we can use this function and take the product over all stars. 
\begin{eqnarray}
p(a|\mbox{data}) &\propto& p(a) \prod_i \int L_i(\tau,\zeta,m) \,  \phi(\zeta) \, \xi(m) \nonumber \\
&\times&  N(a) \, \psi(\tau | a)  \, S_i(\tau,\zeta,m) \diff \tau \diff \zeta \diff m \nonumber
\end{eqnarray}
Now recalling Equation (\ref{likelihood}), we can integrate over mass and use $L(\tau,\zeta)$, and also over metallicity, assuming a flat metallicity distribution.
\begin{eqnarray}
p(a|\mbox{data}) &\propto& p(a) \prod_i \int L_i(\tau) \, N(a) \, \psi(\tau | a) \diff \tau
\label{hierP2}
\end{eqnarray}
We have already computed $L_i(\tau)$ for each star in Section \ref{prob}, therefore we can use these PDFs to constrain the parameter $a$ of the model prior. The likelihood PDFs were calculated under assumed SFH, IMF, and metallicity distribution priors. The assumed IMF and MDF are the same as in Section \ref{prob}. The effect of using a grid of isochrones in log(age) is a flat SFH prior in $\tau$. However, the prior being modeled here is the SFH, which will be a function of $\tau$, therefore we do not need to explicitly remove the prior imposed by the isochrone grid. The model SFH priors also have free parameter(s) $a$, $\psi(\tau|a)$.

We test this hierarchical modeling method on a mock sample of stars generated to have an underlying Gaussian SFH, with a mean age of $\tau=9.0$ and an age dispersion of $\sigma=0.4$, and selected to be only giants using our selection function. We find that the hierarchical modeling accurately recovers the mean age and age dispersion of the simulated sample. The model SFH was found to have a mean age of 9.0 and an age dispersion of 0.37, very similar to the input sample. This demonstrates that the hierarchical modeling method is able to correctly recover the underlying SFH, assuming the model function is a good representation of the true SFH.

We test a few simple models for the SFH of the observed Hipparcos sample, starting with a flat SFH in log(age), $\tau$, given by 
\begin{equation}
\psi(\tau|\tau_{\mbox{\scriptsize min}},\tau_{\mbox{\scriptsize max}}) = 
\begin{cases}
1 & \tau_{\mbox{\scriptsize min}} \leq \tau \leq \tau_{\mbox{\scriptsize max}} \\
0 & \mbox{elsewhere} \nonumber
\end{cases}
\end{equation}
where $\tau_{\mbox{\scriptsize min}}$ and $\tau_{\mbox{\scriptsize max}}$ are the free parameters. To determine the values of $\tau_{\mbox{\scriptsize min}}$ and $\tau_{\mbox{\scriptsize max}}$ we calculate the  probability of the parameter value given the data (Equation (\ref{hierP2})) for a grid of $\tau_{\mbox{\scriptsize min}}$ and $\tau_{\mbox{\scriptsize max}}$ values. We find the most likely parameters to be $\tau_{\mbox{\scriptsize min}}$ of 8.05 and $\tau_{\mbox{\scriptsize max}}$ of 10.1. This is a reasonable range of ages for a sample of solar neighborhood giants, however, the SFH is likely more complicated. We also test models for a linear SFH in age, and a Gaussian SFH in log(age). We find that the linear SFH model cannot recover a likely slope while maintaining a positive SFR across the full age range.

The Gaussian model has the form
\begin{equation}
\psi(\tau|\mu,\sigma)= \frac{1}{\sigma \sqrt{2\pi}} \exp \left( \frac{(\tau - \mu)^2}{2 \sigma^2} \right) \nonumber
\end{equation}
where the mean log(age), $\mu$, and log(age) dispersion, $\sigma$, are the free parameters. The value of these parameters was determined through a similar grid search as the previous model, resulting in a most likely $\mu$ of 9.25 and $\sigma$ of 0.39. This is consistent with the distribution of ages found in Section \ref{prob} using the assumed flat SFH. This is also consistent with the expected age distribution of a sample of giants in the solar neighborhood as shown in Figure \ref{Ntau}. 

Although the Gaussian model finds the overall SFH of the whole sample, there are more individual star PDFs that differ largely from the average fit than would be expected from purely Gaussian wings. This suggests we need a more complex model to account for these outlier stars. 

Motivated by recent work that demonstrates that \ad abundance may correlate more closely with age than [Fe/H] \citep[see e.g. ][and Figure \ref{alphaage} below]{Haywood2013} we test an \ad dependent Gaussian SFH model. The SFH model is applied to a subsample of stars with a single \ad abundance, and most likely parameters determined for each abundance bin. Even with an \ad dependent model we find stars with age PDFs that are significantly inconsistent with the most likely SFH model for the given \ad abundance. We therefore test a uniform+Gaussian \ad dependent SFH model, which consists of a Gaussian SFH plus a constant SFH for some fraction of outlier stars. This model is given by
\begin{equation}
\psi(\tau|\mu,\sigma) = \frac{(1-A)}{\sigma \sqrt{2\pi}} \exp \left( \frac{(\tau - \mu)^2}{2 \sigma^2} \right) + A \times C \nonumber
\end{equation}
where $A$ is the outlier fraction and C is a constant. This allows the single \ad abundance population to be fit by a Gaussian SFH while also allowing for a small fraction of outlier stars. Although the pure Gaussian models are sufficient to examine the mean age of the \ad abundance dependent populations, when determining ages for individual stars, the age PDFs for outlier stars are significantly modified by the SFH prior. The Gaussian+uniform SFH model is needed for the SFH prior used in an empirical Bayesian analysis to determine individual star ages. We find an outlier fraction of $7.5 \%$ is consistent with our sample, however, the value of the outlier fraction has a very small effect on the individual stellar ages. When applied to the whole sample independently of abundance, the Gaussian+uniform model results in a mean $\tau$ of 9.25 and a dispersion of 0.37. This is very similar to the results of the pure Gaussian model suggesting that the constant term does not have a large impact on the fit to the dominant population. 

Our sample has only 64 stars with [\al/M] $> 0.1$, therefore we use single \ad abundance bins independent of [M/H]. The SFH model could be applied to a larger sample of stars for monoabundance populations such as those examined by \citet{Bovy2012a}.

The \ad dependent SFH model provides some useful constraints on the SFH of single \ad abundance populations in the solar neighborhood, see Section \ref{alphasection}. To determine individual stellar ages for the Hipparcos sample, we adopt the \ad dependent, Gaussian+uniform SFH models as the SFH prior in an empirical Bayesian analysis, as was done in Section \ref{prob} assuming a flat SFH in age. For this analysis we construct a full age PDF in log age and again assign a single age to an individual star as the mean of the PDF. The bottom panel of Figure \ref{agedisthip} shows the distribution of ages found for the Hipparcos sample. Using this method we again find a peak in ages at 9.1 and a peak at 9.55. The behavior of this distribution is consistent with the expected distribution for our selection function, however, the exact age of the peaks is slightly shifted from the theoretical prediction in Figure \ref{Ntau} by about 0.1~dex. Again, this offset makes analysis of the underlying population difficult, but is a reasonable distribution given the uncertainty in age. As in Section \ref{hipages}, we use the variance of the individual age PDF as the uncertainty in the age. For this analysis the mean uncertainty is 0.1~dex with a spread of about 0.05~dex. This is a smaller age uncertainty than was found using the uninformed Bayesian analysis in Section \ref{prob}.

\section{Age Trends of The Hipparcos Sample}
\label{agetrends}

\subsection{\ad dependent age distribution}
\label{alphasection}

\begin{figure}[t!]
\centering
\includegraphics[angle=90,width=0.45 \textwidth]{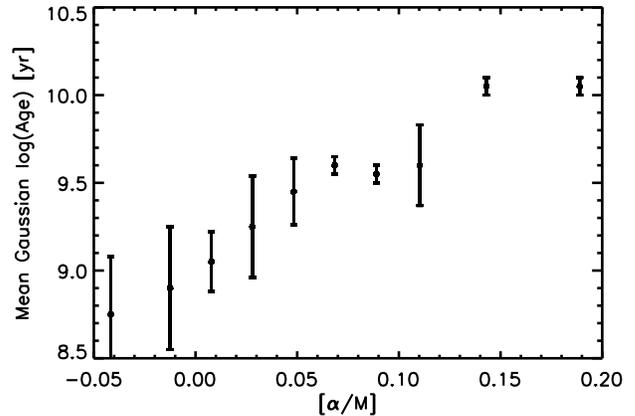}
\caption{ The mean log(age), $\tau$, of the \ad dependent Gaussian+uniform SFH models as a function of \ad abundance bin. The error bars indicate the dispersion of each Gaussian model. The mean age of the models increases continuously with \ad abundance, even at the youngest ($< 1$~Gyr) ages.}
\label{alphagauss}
\end{figure}

The \ad dependent Gaussian+uniform SFH models give a most likely mean $\tau$ and dispersion for each population of stars with a single \ad abundance. The \al \, bins used to fit the models were chosen to have at least 15 stars in each bin, with a minimum bin width of 0.02~dex in [\al/M]. Figure \ref{alphagauss} shows the mean $\tau$ of the Gaussian SFH model as a function of \ad abundance. The $\tau$ dispersion of each model is indicated by the error bars. The mean age increases strongly with \ad abundance, even at solar \ad abundance and young ages. Recent work from the Gaia-ESO survey has shown a large spread in \ad abundance at a given age with a very shallow increase in \ad abundance \citep{Bergemann2014} . Other studies in the solar neighborhood have found little trend in \ad abundance at young ages and a steep increase in \ad abundance with age above about 8~Gyr \citep[e.g.][]{Ramirez2013, Haywood2013, Bensby2014}. The hierarchical modeling shows that there is a correlation between \ad abundance and age across the full abundance range. We note also that the dispersion in the SFH model is smaller for the more \ad rich populations, and larger for the solar-\al \, populations. This suggests that while the overall [\al/M] decreased with time in the Galactic disk, some solar \ad abundance stars formed early on, while the formation of \ad enhanced stars was limited to early times.

\begin{figure}[t!]
\centering
\includegraphics[angle=90,width=0.45 \textwidth]{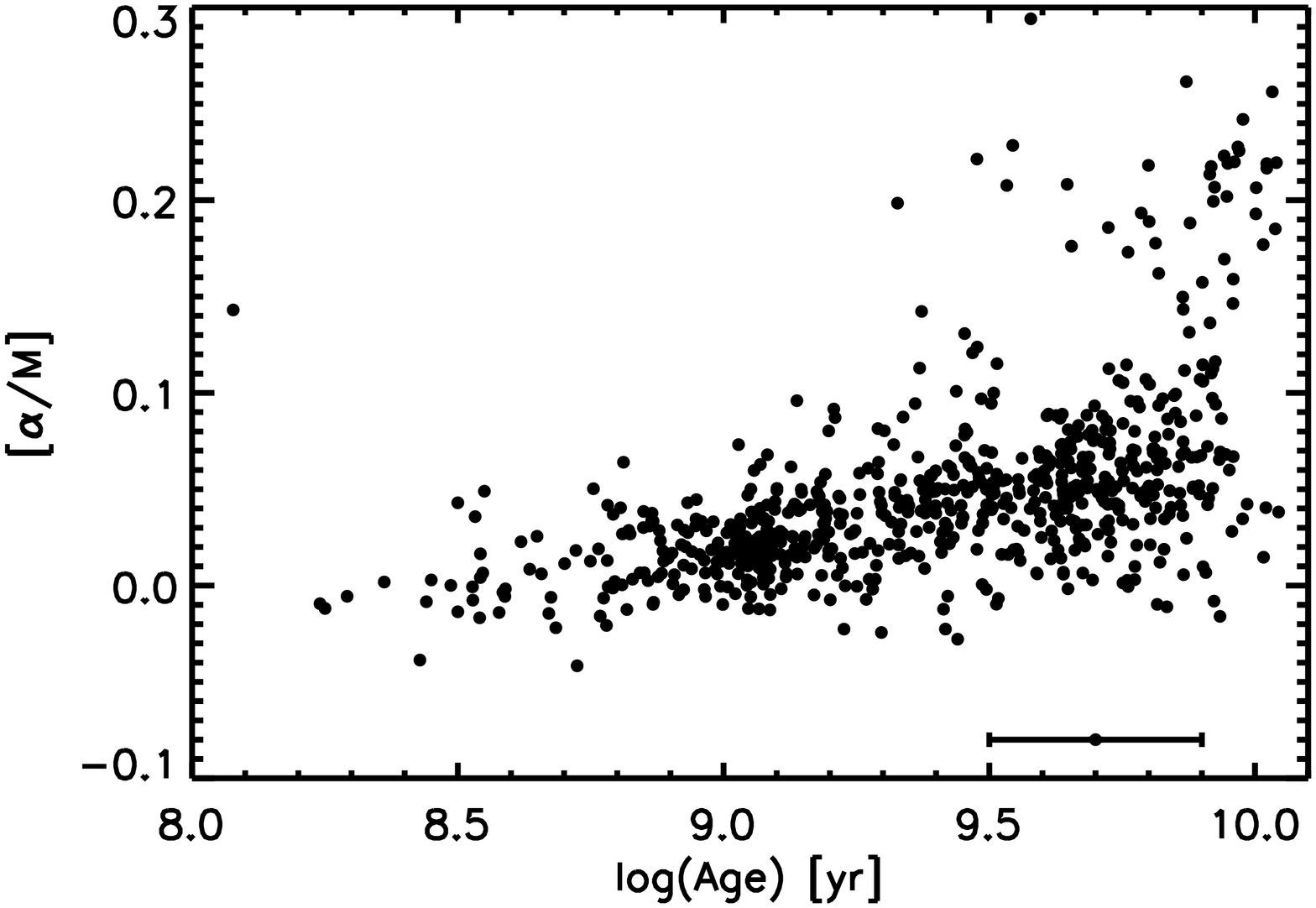}
\includegraphics[angle=90,width=0.45 \textwidth]{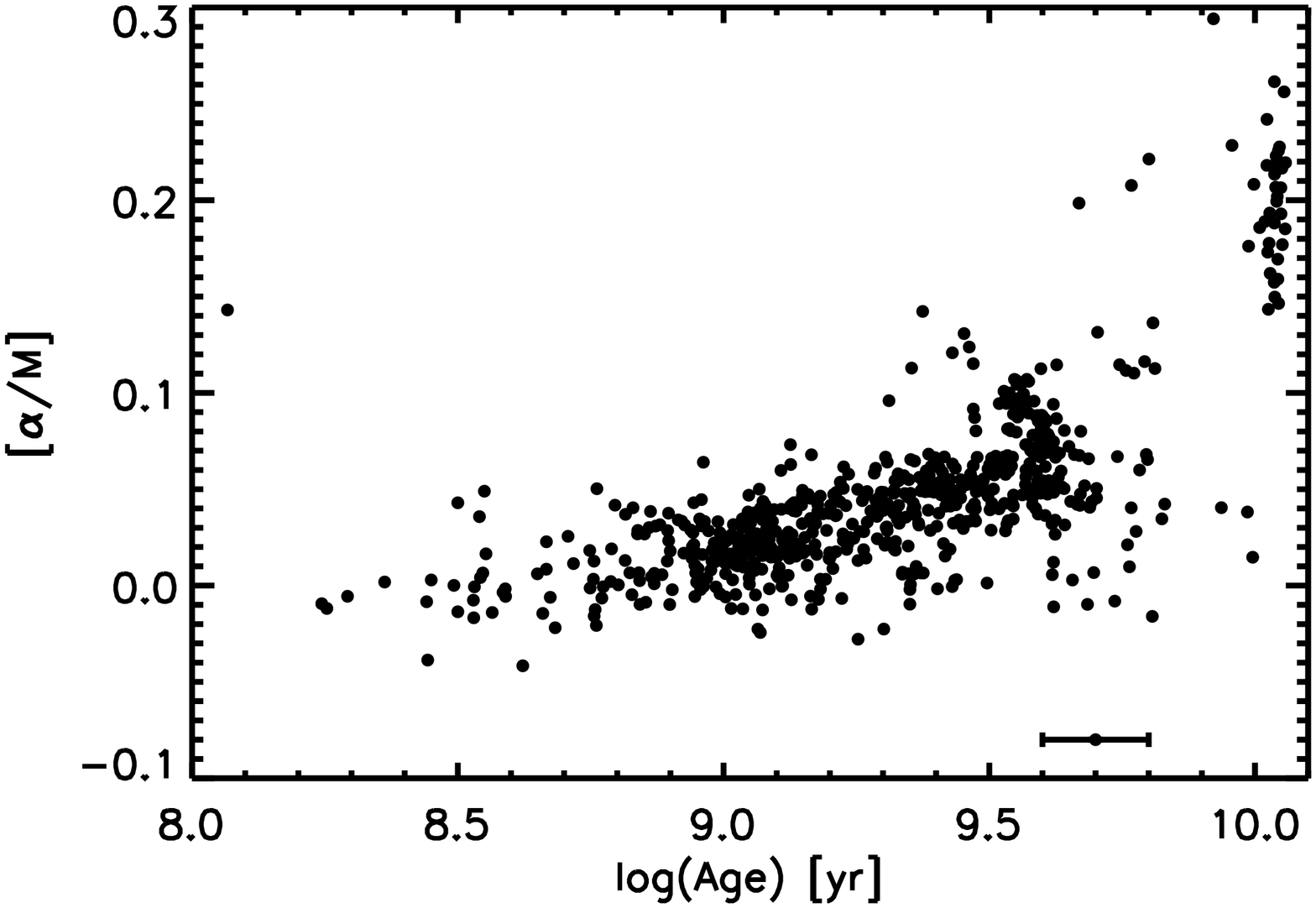}
\caption{ [\al/M] as a function of age. The top panel uses ages found using the initial Bayesian method described in Section \ref{prob}. The bottom panel uses ages found using the empirical Bayesian method described in Section \ref{hierarchical}. As expected from the \ad dependent SFH prior, the relation is much tighter using ages determined through the empirical Bayesian analysis. However, there are a few young, \ad rich stars and older, \ad poor stars. Typical uncertainties are shown. 
}
\label{alphaage}
\end{figure}

This analysis is useful for general trends in the disk populations with time, however, we would also like to examine how the individual stars deviate from that overall trend. The ages for individual stars determined using the \ad dependent Gaussian+uniform model SFH as a prior in an empirical Bayesian analysis are shown as a function of stellar \ad abundance in the bottom panel of Figure \ref{alphaage}. The tight relation between \ad abundance and age dictated by the SFH model is recovered. However, there is still some spread in \al \, at all ages. Using the Bayesian ages, top panel of Figure \ref{alphaage}, the relation between [\al/M] and age is still present, confirming the motivation to use an \ad dependent model SFH. The spread in \al \, is much larger for the Bayesian ages than for the empirical Bayesian ages, particularly older than 1~Gyr. Although the trend between [\al/M] and age is consistent with previous work, our sample does not have stars past 11~Gyr as is seen in some studies of the Galactic disk \citep[e.g. ][]{Haywood2013, Bergemann2014, Bensby2014}. This is likely due to our method of taking the mean of the age PDF and the isochrone grid edge of 12.6~Gyr, creating a bias against the oldest stars, as discussed in Section \ref{prob}.

Although the empirical Bayesian ages include an \ad dependent prior, there are still a few outlier stars, suggesting that these stars are robustly very young or very old for their \ad abundance. Recently, young \ad rich stars have been found using asteroseimology in the Kepler \citep{Martig2015} and CoRoT \citep{Chiappini2015a} fields. We find six stars in our sample that are \ad rich ([\al/M] $> 0.13$) and strongly constrained to be younger than $\tau = 9.75 \sim 6$~Gyr. We also find seven stars with [\al/M] $< 0.1$ and older than $\tau = 9.75 \sim 6$~Gyr.

\begin{figure}[t!]
\centering
\includegraphics[angle=90,width=0.45 \textwidth]{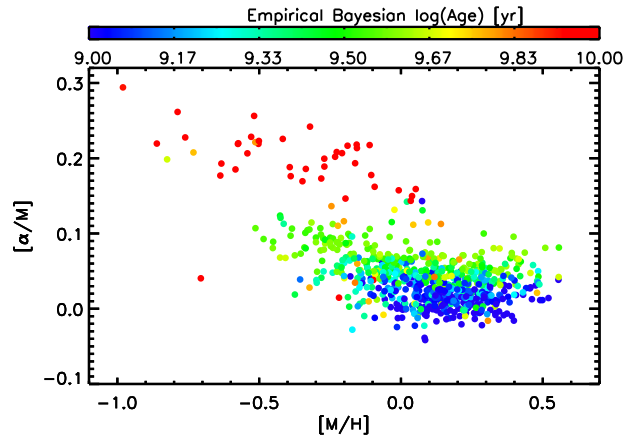}
\caption{ The [\al/M] vs [M/H] distribution for our sample, color-coded by age. There is a separation of high and low \ad abundance stars with metallicity. The low-\al \, stars have a ``banana" distribution as seen in recent studies of the Milky Way disk. The \ad rich stars are older than the \ad poor stars, in agreement with recent studies. We note the presence of 16 \ad rich stars that are younger than $\tau = 9.75 \sim 6$~Gyr.}
\label{alpha}
\end{figure}

Previous work has relied on the [\al/Fe] vs [Fe/H] plane to study the evolution of the thick and thin disk populations \citep[e.g. ][]{Fuhrmann1998, Prochaska2000a, Reddy2006, Haywood2013, Bensby2014, Hayden2015}, finding the thick disk to be older and \ad enhanced, and the thin disk to be younger with solar \ad abundance. We examine the [\al/M] vs. [M/H] relation of this sample, shown in Figure \ref{alpha}, with the added dimension of age, indicated by the color. Our sample shows a separation of low-\al \, and high-\al \, abundance populations as is typical of the local disk. The high-\al \, track is older than the low-\al \, track, however, the low-\al \, track does contain 16 stars older than 6~Gyr. The ages of the high-\al \, track are consistent with the age of the thick disk determined from the white dwarf luminosity function \citep[see e.g. ][]{Leggett1998, Reid2005}.
The low-\al \, population shows a ``banana"-shaped distribution as is seen by \citet{Adibekyan2012} in their sample of FGK stars, \citet{Nidever2014} in a sample of RC stars observed by the APOGEE survey, and \citet{Bensby2014} in the Ti abundances of local FG dwarfs and subgiants. 

\subsection{Age-metallicity relation}
\label{metalsection}

\begin{figure}[t]
\centering
\includegraphics[angle=90,width=0.45 \textwidth]{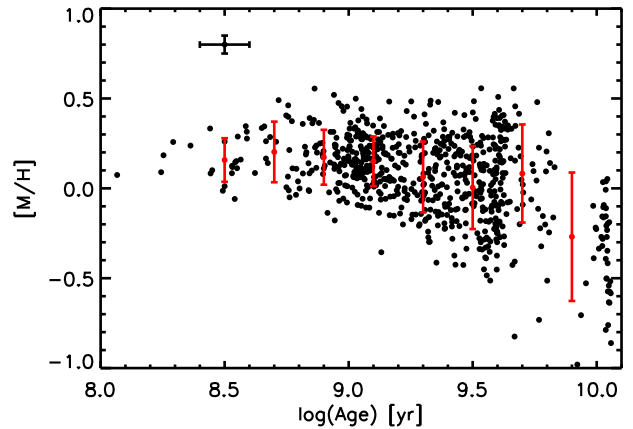}
\caption{ Metallicity as a function of age for giants in the solar neighborhood. We find a flat age-metallicity relation at younger ages with decreasing metallicity around $\tau$ of $9.75 \sim 6$~Gyr. A representative age uncertainty is given in the top left. The red points indicate the mean metallicity in age bins of 0.2~dex and the error bars indicate the standard deviation in metallicity in each bin. We note that the scatter in metallicity is smaller for stars with ages $< 1$~Gyr.}
\label{age-metal}
\end{figure}

The presence of an age-metallicity relation is a basic prediction of simple stellar and galactic evolution models; as stars evolve they enrich the galaxy with new metals so that successive generations of stars are more metal rich. Recent spectroscopic studies of stars in the solar neighborhood have found a deviation from this expected age-metallicity relation. \citet{Edvardsson1993}, \citet{Haywood2013}, \citet{Bensby2014}, and \citet{Bergemann2014} all find a $\sim$ 1.0~dex spread in metallicity at all ages, with a flat relation at ages younger than 8~Gyr and the predicted decrease in metallicity with increased age present only in stars older than 8~Gyr. The age-metallicity relation derived from our sample is shown in Figure \ref{age-metal}. In agreement with recent studies, we find a weak age-metallicity relation at younger ages with a steeper decrease in metallicity at around 6~Gyr and a spread in metallicity across all ages. Our results agree with previous work to suggest that the chemical evolution of the Milky Way disk was more complicated than a simple closed box model. 

The red points in Figure \ref{age-metal} indicate the mean [M/H] and standard deviation in [M/H] for age bins of 0.2~dex. This is consistent with a very shallow age-metallicity relation below $\sim 6$~Gyr. More interestingly, we note that the scatter in the metallicity is smaller for stars with ages $< 1$~Gyr. This could reflect the effects of radial migration on the MDF of the solar neighborhood. Young stars that formed near the solar neighborhood should have a similar metallicity, while star than formed elsewhere could have a different metallicity. These older stars have had time to migrate to the solar neighborhood, causing a spread in the MDF of older stars.

\subsection{Kinematics}

\begin{figure*}[t!]
\centering
\includegraphics[angle=90,width=0.45 \textwidth]{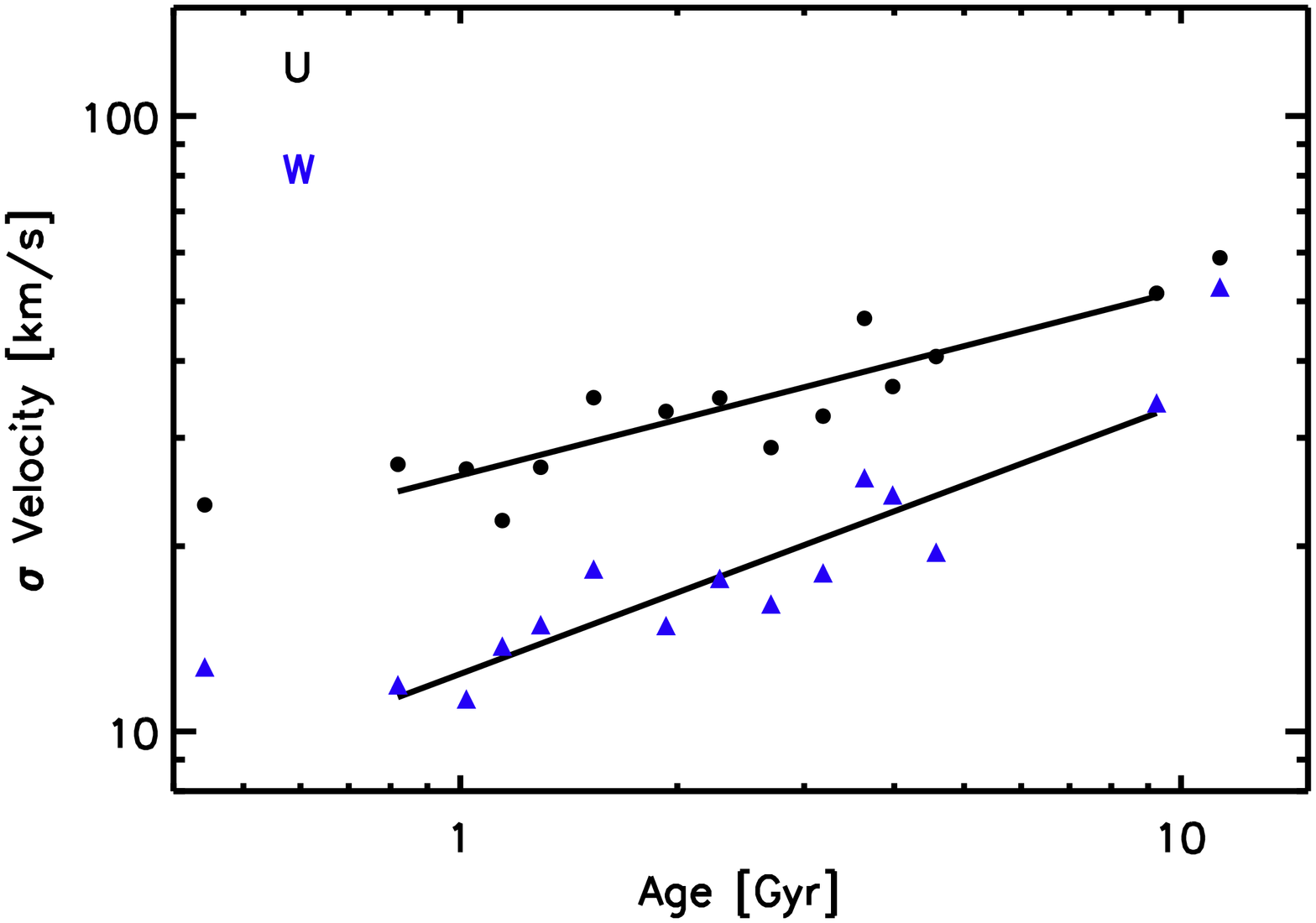}
\includegraphics[angle=90,width=0.45 \textwidth]{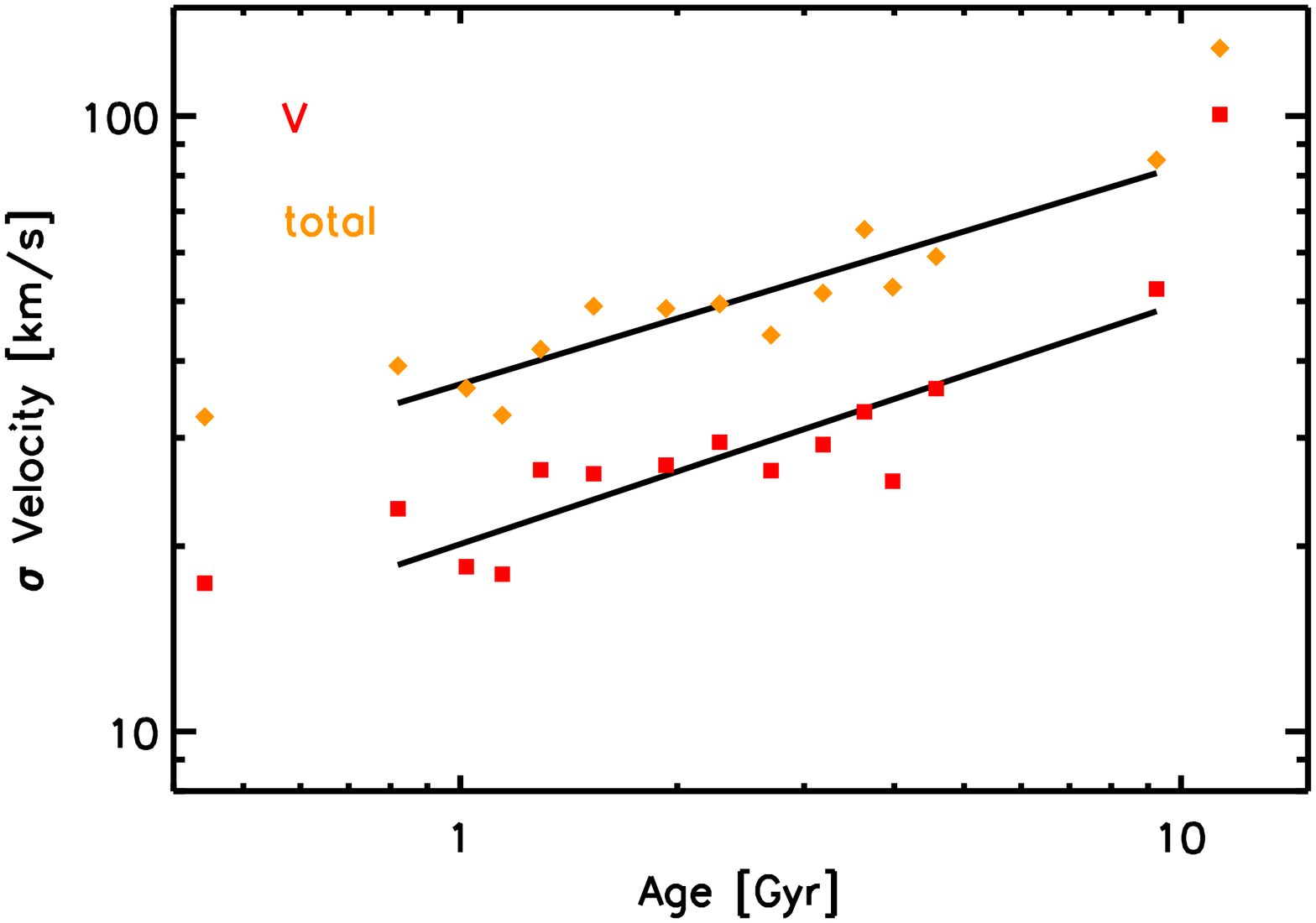}
\caption{ The velocity dispersion as a function of age. On the left: U and W velocity components and on the right: V and total velocity for our sample. Each bin contains equal numbers of stars (50 stars each). This plot is comparable to Figure 7 of \citet{Holmberg2009}. The line indicates the best fit power law to the data. The highest and lowest age bins are excluded from the fit. The indices of the power law for the U, V, W, and total velocities are 0.30, 0.39, 0.44, and 0.36, respectively. The fits are normalized to 26.0, 20.2, 12.4, and 36.7~km~s$^{-1}$ at 1~Gyr for the U, V, W, and total velocity components, respectively.}
\label{vdisp}
\end{figure*}

As full kinematic information is available for this sample, we examine trends in kinematics with age. We find that the spread in velocity increases with age in all velocity components, in agreement with the GCS results from \citet{Holmberg2009}. Figure \ref{vdisp} shows the resulting velocity dispersion as a function of age with this sample. In this figure, each age bin contains equal numbers of stars, 50 stars each. We find that the velocity dispersion increases with age for all velocity components. The line indicates the best fit power law to the data. The highest and lowest age bin is excluded from the fit. The indices of the power law for the U, V, W, and total velocities are 0.30, 0.39, 0.44, and 0.36, respectively. The fits are normalized to 26.0, 20.2, 12.4, and 36.7~km~s$^{-1}$ at 1~Gyr for the U, V, W, and total velocity components, respectively. The spread around the exponential fit is 4, 3, 4, and 5~km~s$^{-1}$ for the U, V, W, and total velocity components, respectively. Visually, the trends and magnitude of the velocity dispersion with age are consistent with the improved GCS results \citep{Holmberg2009}, and the power law indices agree within 0.1 for all velocity components.

\section{Conclusions}
\label{conclusion}

We present a method of determining ages to within 0.15~dex for red giant stars with accurately measured distances and high resolution spectroscopic parameters. Due to the probabilistic approach and the large uncertainty associated with isochrone matching on the giant branch this method is recommended for a statistical examination of stellar populations. We test the Bayesian and hierarchical modeling methods on a mock sample of stars to test the accuracy of the ages. We then apply this method to a sample of 705 local stars with distances measured to within 10 \% by the Hipparcos mission and observed with the 1m+APOGEE capability. 

Due to the large uncertainties in the mass calculated from the luminosity, effective temperature, and surface gravity, we find that a direct mass-age relation results in an age uncertainty of 0.3~dex. In order to reduce this uncertainty in age without better mass measurements, a probabilistic isochrone matching method is required. Using a mock sample of stars we find that a Bayesian analysis assuming a Chabrier lognormal IMF and a flat SFH is able to recover the true age of giants to within 0.2~dex. For this analysis, [Fe/H], $T_{\mbox{\small eff}}$, $M_{V_T}$, and $\log g$ are the best measured parameters to match to the isochrones. When applied to a sample of solar neighborhood giant stars, the resulting age distribution is consistent with the age distribution expected for our selection function.

In order to more accurately model the SFH, we apply a hierarchical modeling method. The hierarchical modeling approach uses the full age PDF of the data to constrain the parameters of a model SFH. The model SFH can be fit to the entire sample or to subsamples based on abundance or location. This method is powerful because it allows the data to inform the SFH and the SFH can be tuned for specific subsamples. In this paper, we fit the model SFH to subsamples of single \ad abundance stars and find an \ad dependent Gaussian+uniform SFH model represents the data well. From this analysis we find the mean log(age) of single \ad abundance populations increases with \ad abundance, and the dispersion in the SFH is larger for solar [\al/M] populations that for \ad enhanced populations. Although previous work has found a flat \ad age relation below $\sim 7$~Gyr, using the hierarchical modeling we find a relation does exist at younger ages. In agreement with recent studies of the solar neighborhood, we find the \ad rich track in [\al/M] vs [M/H] space is older than the low-\al \, track. In future work we plan to test a model SFH fit to monoabundance populations across the Galactic disk. 

Using this model SFH as the prior in an empirical Bayesian analysis we determine individual stellar ages and examine the trends in abundance with age for the solar neighborhood.  With the individual stellar ages we do note the presence of six \ad rich stars with ages below 6~Gyr as seen by \citet{Martig2015} and \citet{Chiappini2015a}. We also find seven \ad poor stars with ages above 6~Gyr.

As previously found, the age-metallicity relation of our sample is flat at young ages and decreases in metallicity around 6~Gyr. Although there is a spread in [M/H] at all ages, we find that the spread for stars younger than $\sim 1$~Gyr is smaller than for older stars. This could be a result of radial migration as older stars born at different Galactic radii migrate to the solar neighborhood and broaden the MDF.

The velocity dispersion in this sample increases with age in all velocity components, in agreement with the improved GSC results \citep{Holmberg2009}.

The trends with age of this sample agree with other work using stars from the solar neighborhood. We find this method of determining ages of red giant stars to be applicable to large samples of giants for which accurate distance measurements and high resolution spectroscopy is available.

\acknowledgments

Funding for SDSS-III has been provided by the Alfred P. Sloan Foundation, the Participating Institutions, the National Science Foundation, and the U.S. Department of Energy Office of Science. The SDSS-III web site is http://www.sdss3.org/.
SDSS-III is managed by the Astrophysical Research Consortium for the Participating Institutions of the SDSS-III Collaboration including the University of Arizona, the Brazilian Participation Group, Brookhaven National Laboratory, Carnegie Mellon University, University of Florida, the French Participation Group, the German Participation Group, Harvard University, the Instituto de Astrofisica de Canarias, the Michigan State/Notre Dame/JINA Participation Group, Johns Hopkins University, Lawrence Berkeley National Laboratory, Max Planck Institute for Astrophysics, Max Planck Institute for Extraterrestrial Physics, New Mexico State University, New York University, Ohio State University, Pennsylvania State University, University of Portsmouth, Princeton University, the Spanish Participation Group, University of Tokyo, University of Utah, Vanderbilt University, University of Virginia, University of Washington, and Yale University.

J.B. thanks the Natural Sciences and Engineering Research Council of Canada for financial support of this project.

\bibliographystyle{apj}
\bibliography{hip_age_astroph}

\begin{thebibliography}{57}
\expandafter\ifx\csname natexlab\endcsname\relax\def\natexlab#1{#1}\fi

\bibitem[{Adibekyan {et~al.}(2012)Adibekyan, Sousa, Santos, {Delgado Mena},
  {Gonz\'{a}lez Hern\'{a}ndez}, Israelian, Mayor, \&
  Khachatryan}]{Adibekyan2012}
Adibekyan, V.~Z., Sousa, S.~G., Santos, N.~C., {et~al.} 2012, A \& A, 545, A32

\bibitem[{{Alam} {et~al.}(2015){Alam}, {Albareti}, {Allende Prieto}, {Anders},
  {Anderson}, {Anderton}, {Andrews}, {Armengaud}, {Aubourg}, {Bailey}, \&
  et~al.}]{Alam2015}
{Alam}, S., {Albareti}, F.~D., {Allende Prieto}, C., {et~al.} 2015, ApJS, 219,
  12

\bibitem[{Anders {et~al.}(2015)Anders, Chiappini, Rodrigues, Miglio,
  Montalb\'{a}n, Mosser, Girardi, Valentini, Noels, Morel, Johnson, Schultheis,
  Baudin, Peralta, Hekker, \& Theme\ss~l}]{Anders2015}
Anders, F., Chiappini, C., Rodrigues, T.~S., {et~al.} 2015, in prep

\bibitem[{{Anderson} \& {Francis}(2012)}]{Anderson2012}
{Anderson}, E., \& {Francis}, C. 2012, VizieR Online Data Catalog, 5137, 0

\bibitem[{{Bahcall} {et~al.}(1995){Bahcall}, {Pinsonneault}, \&
  {Wasserburg}}]{Bahcall1995}
{Bahcall}, J.~N., {Pinsonneault}, M.~H., \& {Wasserburg}, G.~J. 1995, Reviews
  of Modern Physics, 67, 781

\bibitem[{{Bensby} {et~al.}(2014){Bensby}, {Feltzing}, \& {Oey}}]{Bensby2014}
{Bensby}, T., {Feltzing}, S., \& {Oey}, M.~S. 2014, A \& A, 562, A71

\bibitem[{Bergemann {et~al.}(2014)Bergemann, Ruchti, Serenelli, Feltzing,
  Alves-Brito, Asplund, Bensby, Gruiters, Heiter, Hourihane, Korn, Lind,
  Marino, Jofre, Nordlander, Ryde, Worley, Gilmore, Randich, Ferguson,
  Jeffries, Micela, Negueruela, Prusti, Rix, Vallenari, Alfaro, {Allende
  Prieto}, Bragaglia, Koposov, Lanzafame, Pancino, Recio-Blanco, Smiljanic,
  Walton, Costado, Franciosini, Hill, Lardo, de~Laverny, Magrini, Maiorca,
  Masseron, Morbidelli, Sacco, Kordopatis, \&
  Tautvai\v{s}ienė}]{Bergemann2014}
Bergemann, M., Ruchti, G.~R., Serenelli, A., {et~al.} 2014, A \& A, 565, A89

\bibitem[{Boeche {et~al.}(2013)Boeche, Chiappini, Minchev, Williams, Steinmetz,
  Sharma, Kordopatis, Bland-Hawthorn, Bienaym\'{e}, Gibson, Gilmore, Grebel,
  Helmi, Munari, Navarro, Parker, Reid, Seabroke, Siebert, Siviero, Watson,
  Wyse, \& Zwitter}]{Boeche2013}
Boeche, C., Chiappini, C., Minchev, I., {et~al.} 2013, A \& A, 553, A19

\bibitem[{{Bovy} {et~al.}(2012){Bovy}, {Rix}, {Liu}, {Hogg}, {Beers}, \&
  {Lee}}]{Bovy2012a}
{Bovy}, J., {Rix}, H.-W., {Liu}, C., {et~al.} 2012, ApJ, 753, 148

\bibitem[{Bovy {et~al.}(2014)Bovy, Nidever, Rix, Girardi, Zasowski, Chojnowski,
  Holtzman, Epstein, Frinchaboy, Hayden, Rodrigues, Majewski, Johnson,
  Pinsonneault, Stello, {Allende Prieto}, Andrews, Basu, Beers, Bizyaev,
  Burton, Chaplin, Cunha, Elsworth, Garc\'{\i}a, Garc\'{\i}a-Herńandez,
  {Garc\'{\i}a P\'{e}rez}, Hearty, Hekker, Kallinger, Kinemuchi, Koesterke,
  M\'{e}sz\'{a}ros, Mosser, O'Connell, Oravetz, Pan, Robin, Schiavon,
  Schneider, Schultheis, Serenelli, Shetrone, {Silva Aguirre}, Simmons,
  Skrutskie, Smith, Stassun, Weinberg, Wilson, \& Zamora}]{Bovy2014a}
Bovy, J., Nidever, D.~L., Rix, H.-W., {et~al.} 2014, ApJ, 790, 127

\bibitem[{Bressan {et~al.}(2012)Bressan, Marigo, Girardi, Salasnich, {Dal
  Cero}, Rubele, \& Nanni}]{Bressan2012}
Bressan, A., Marigo, P., Girardi, L., {et~al.} 2012, MNRAS, 427, 127

\bibitem[{Chabrier(2001)}]{Chabrier2001}
Chabrier, G. 2001, ApJ, 554, 1274

\bibitem[{Chang {et~al.}(1999)Chang, Hou, Shu, \& Fu}]{Chang1999}
Chang, R.~X., Hou, J.~L., Shu, C.~G., \& Fu, C.~Q. 1999, A \& A, 350, 38

\bibitem[{Chiappini {et~al.}(1997)Chiappini, Matteucci, \&
  Gratton}]{Chiappini1997}
Chiappini, C., Matteucci, F., \& Gratton, R. 1997, ApJ, 477, 765

\bibitem[{Chiappini {et~al.}(2001)Chiappini, Matteucci, \&
  Romano}]{Chiappini2001}
Chiappini, C., Matteucci, F., \& Romano, D. 2001, ApJ, 554, 1044

\bibitem[{Chiappini {et~al.}(2015)Chiappini, Anders, Rodrigues, Miglio,
  Montalban, Mosser, Girardi, Valentini, Noels, Morel, Minchev, Steinmetz,
  Santiago, Schultheis, Martig, da~Costa, Maia, Prieto, Peralta, Hekker,
  Theme\ss~l, Kallinger, Garcia, Mathur, Baudin, Beers, Cunha, Harding,
  Holtzman, Majewski, Meszaros, Nidever, Pan, Schiavon, Shetrone, Schneider, \&
  Stassun}]{Chiappini2015a}
Chiappini, C., Anders, F., Rodrigues, T.~S., {et~al.} 2015, A \& A, 576, L12

\bibitem[{Edvardsson {et~al.}(1993)Edvardsson, Andersen, Gustafsson, Lambert,
  Nissen, \& Tomkin}]{Edvardsson1993}
Edvardsson, B., Andersen, J., Gustafsson, B., {et~al.} 1993, A \& AS, 102

\bibitem[{Eisenstein {et~al.}(2011)Eisenstein, Weinberg, Agol, Aihara, {Allende
  Prieto}, Anderson, Arns, Aubourg, Bailey, Balbinot, Barkhouser, Beers,
  Berlind, Bickerton, Bizyaev, Blanton, Bochanski, Bolton, Bosman, Bovy,
  Brandt, Breslauer, Brewington, Brinkmann, Brown, Brownstein, Burger, Busca,
  Campbell, Cargile, Carithers, Carlberg, Carr, Chang, Chen, Chiappini,
  Comparat, Connolly, Cortes, Croft, Cunha, da~Costa, Davenport, Dawson, {De
  Lee}, {Porto de Mello}, de~Simoni, Dean, Dhital, Ealet, Ebelke, Edmondson,
  Eiting, Escoffier, Esposito, Evans, Fan, {Femen\'{\i}a Castell\'{a}}, {Dutra
  Ferreira}, Fitzgerald, Fleming, Font-Ribera, Ford, Frinchaboy, {Garc\'{\i}a
  P\'{e}rez}, Gaudi, Ge, Ghezzi, Gillespie, Gilmore, Girardi, Gott, Gould,
  Grebel, Gunn, Hamilton, Harding, Harris, Hawley, Hearty, Hennawi,
  {Gonz\'{a}lez Hern\'{a}ndez}, Ho, Hogg, Holtzman, Honscheid, Inada, Ivans,
  Jiang, Jiang, Johnson, Jordan, Jordan, Kauffmann, Kazin, Kirkby, Klaene,
  Knapp, Kneib, Kochanek, Koesterke, Kollmeier, Kron, Lampeitl, Lang, Lawler,
  {Le Goff}, Lee, Lee, Leisenring, Lin, Liu, Long, Loomis, Lucatello, Lundgren,
  Lupton, Ma, Ma, MacDonald, Mack, Mahadevan, Maia, Majewski, Makler,
  Malanushenko, Malanushenko, Mandelbaum, Maraston, Margala, Maseman, Masters,
  McBride, McDonald, McGreer, McMahon, {Mena Requejo}, M\'{e}nard,
  Miralda-Escud\'{e}, Morrison, Mullally, Muna, Murayama, Myers, Naugle, Neto,
  Nguyen, Nichol, Nidever, O’Connell, Ogando, Olmstead, Oravetz, Padmanabhan,
  Paegert, Palanque-Delabrouille, Pan, Pandey, Parejko, P\^{a}ris, Pellegrini,
  Pepper, Percival, Petitjean, Pfaffenberger, Pforr, Phleps, Pichon, Pieri,
  Prada, Price-Whelan, Raddick, Ramos, Reid, Reyle, Rich, Richards, Rieke,
  Rieke, Rix, Robin, Rocha-Pinto, Rockosi, Roe, Rollinde, Ross, Ross, Rossetto,
  S\'{a}nchez, Santiago, Sayres, Schiavon, Schlegel, Schlesinger, Schmidt,
  Schneider, Sellgren, Shelden, Sheldon, Shetrone, Shu, Silverman, Simmerer,
  Simmons, Sivarani, Skrutskie, Slosar, Smee, Smith, Snedden, Stassun, Steele,
  Steinmetz, Stockett, Stollberg, Strauss, Szalay, Tanaka, Thakar, Thomas,
  Tinker, Tofflemire, Tojeiro, Tremonti, {Vargas Maga\~{n}a}, Verde, Vogt,
  Wake, Wan, Wang, Weaver, White, White, Wilson, Wisniewski, Wood-Vasey, Yanny,
  Yasuda, Y\`{e}che, York, Young, Zasowski, Zehavi, \& Zhao}]{Eisenstein2011}
Eisenstein, D.~J., Weinberg, D.~H., Agol, E., {et~al.} 2011, AJ, 142, 72

\bibitem[{{Freeman}(2012)}]{Freeman2012}
{Freeman}, K.~C. 2012, in Astronomical Society of the Pacific Conference
  Series, Vol. 458, Galactic Archaeology: Near-Field Cosmology and the
  Formation of the Milky Way, ed. W.~{Aoki}, M.~{Ishigaki}, T.~{Suda},
  T.~{Tsujimoto}, \& N.~{Arimoto}, 393

\bibitem[{Fuhrmann(1998)}]{Fuhrmann1998}
Fuhrmann, K. 1998, A \& A, 183, 161

\bibitem[{{Garc{\'{\i}}a P{\'e}rez} {et~al.}(2015){Garc{\'{\i}}a P{\'e}rez},
  {Allende Prieto}, {Holtzman}, {Shetrone}, {M{\'e}sz{\'a}ros}, {Bizyaev},
  {Carrera}, {Cunha}, {Garc{\'{\i}}a-Hern{\'a}ndez}, {Johnson}, {Majewski},
  {Nidever}, {Schiavon}, {Shane}, {Smith}, {Sobeck}, {Troup}, {Zamora}, {Bovy},
  {Eisenstein}, {Feuillet}, {Frinchaboy}, {Hayden}, {Hearty}, {Nguyen},
  {O'Connell}, {Pinsonneault}, {Weinberg}, {Wilson}, \&
  {Zasowski}}]{Garcia2015}
{Garc{\'{\i}}a P{\'e}rez}, A.~E., {Allende Prieto}, C., {Holtzman}, J.~A.,
  {et~al.} 2015, AJ, submitted, arXiv:1510.07635

\bibitem[{{Gilmore} {et~al.}(2012){Gilmore}, {Randich}, {Asplund}, {Binney},
  {Bonifacio}, {Drew}, {Feltzing}, {Ferguson}, {Jeffries}, {Micela},
  {Negueruela}, {Prusti}, {Rix}, {Vallenari}, {Alfaro}, {Allende-Prieto},
  {Babusiaux}, {Bensby}, {Blomme}, {Bragaglia}, {Flaccomio}, {Fran{\c c}ois},
  {Irwin}, {Koposov}, {Korn}, {Lanzafame}, {Pancino}, {Paunzen},
  {Recio-Blanco}, {Sacco}, {Smiljanic}, {Van Eck}, \& {Walton}}]{Gilmore2012}
{Gilmore}, G., {Randich}, S., {Asplund}, M., {et~al.} 2012, The Messenger, 147,
  25

\bibitem[{{Gonz\'{a}lez Hern\'{a}ndez} \&
  Bonifacio(2009)}]{GonzalezHernandez2009}
{Gonz\'{a}lez Hern\'{a}ndez}, J.~I., \& Bonifacio, P. 2009, A \& A, 497, 497

\bibitem[{Gunn {et~al.}(2006)Gunn, Siegmund, Mannery, Owen, Hull, Leger, Carey,
  Knapp, York, Boroski, Kent, Lupton, Rockosi, Evans, Waddell, Anderson, Annis,
  Barentine, Bartoszek, Bastian, Bracker, Brewington, Briegel, Brinkmann,
  Brown, Carr, Czarapata, Drennan, Dombeck, Federwitz, Gillespie, Gonzales,
  Hansen, Harvanek, Hayes, Jordan, Kinney, Klaene, Kleinman, Kron, Kresinski,
  Lee, Limmongkol, Lindenmeyer, Long, Loomis, McGehee, Mantsch, {Neilsen, Jr.},
  Neswold, Newman, Nitta, {Peoples, Jr.}, Pier, Prieto, Prosapio, Rivetta,
  Schneider, Snedden, \& Wang}]{Gunn2006}
Gunn, J.~E., Siegmund, W.~a., Mannery, E.~J., {et~al.} 2006, AJ, 131, 2332

\bibitem[{Hayden {et~al.}(2014)Hayden, Holtzman, Bovy, Majewski, Johnson,
  {Allende Prieto}, Beers, Cunha, Frinchaboy, {Garc\'{\i}a P\'{e}rez}, Girardi,
  Hearty, Lee, Nidever, Schiavon, Schlesinger, Schneider, Schultheis, Shetrone,
  Smith, Zasowski, Bizyaev, Feuillet, Hasselquist, Kinemuchi, Malanushenko,
  Malanushenko, O'Connell, Pan, \& Stassun}]{Hayden2014}
Hayden, M.~R., Holtzman, J.~a., Bovy, J., {et~al.} 2014, AJ, 147, 116

\bibitem[{Hayden {et~al.}(2015)Hayden, Bovy, Holtzman, Nidever, Bird, Weinberg,
  Andrews, Majewski, Prieto, Anders, Beers, Bizyaev, Chiappini, Cunha,
  Frinchaboy, Garc\'{\i}a-Herńandez, {Garc\'{\i}a P\'{e}rez}, Girardi,
  Harding, Hearty, Johnson, M\'{e}sz\'{a}ros, Minchev, O’Connell, Pan, Robin,
  Schiavon, Schneider, Schultheis, Shetrone, Skrutskie, Steinmetz, Smith,
  Wilson, Zamora, \& Zasowski}]{Hayden2015}
Hayden, M.~R., Bovy, J., Holtzman, J.~a., {et~al.} 2015, ApJ, 808, 132

\bibitem[{Haywood {et~al.}(2013)Haywood, {Di Matteo}, Lehnert, Katz, \&
  G\'{o}mez}]{Haywood2013}
Haywood, M., {Di Matteo}, P., Lehnert, M.~D., Katz, D., \& G\'{o}mez, A. 2013,
  A \& A, 560, A109

\bibitem[{Holmberg {et~al.}(2009)Holmberg, Nordstr\"{o}m, \&
  Andersen}]{Holmberg2009}
Holmberg, J., Nordstr\"{o}m, B., \& Andersen, J. 2009, A \& A, 501, 941

\bibitem[{{Holtzman} {et~al.}(2010){Holtzman}, {Harrison}, \&
  {Coughlin}}]{Holtzman2010}
{Holtzman}, J.~A., {Harrison}, T.~E., \& {Coughlin}, J.~L. 2010, Advances in
  Astronomy, 2010

\bibitem[{{Holtzman} {et~al.}(2015){Holtzman}, {Shetrone}, {Johnson}, {Allende
  Prieto}, {Anders}, {Andrews}, {Beers}, {Bizyaev}, {Blanton}, {Bovy},
  {Carrera}, {Chojnowski}, {Cunha}, {Eisenstein}, {Feuillet}, {Frinchaboy},
  {Galbraith-Frew}, {Garc{\'{\i}}a P{\'e}rez}, {Garc{\'{\i}}a-Hern{\'a}ndez},
  {Hasselquist}, {Hayden}, {Hearty}, {Ivans}, {Majewski}, {Martell},
  {Meszaros}, {Muna}, {Nidever}, {Nguyen}, {O'Connell}, {Pan}, {Pinsonneault},
  {Robin}, {Schiavon}, {Shane}, {Sobeck}, {Smith}, {Troup}, {Weinberg},
  {Wilson}, {Wood-Vasey}, {Zamora}, \& {Zasowski}}]{Holtzman2015}
{Holtzman}, J.~A., {Shetrone}, M., {Johnson}, J.~A., {et~al.} 2015, AJ, 150,
  148

\bibitem[{{J{\o}rgensen} \& {Lindegren}(2005)}]{Jorgensen2005}
{J{\o}rgensen}, B.~R., \& {Lindegren}, L. 2005, A \& A, 436, 127

\bibitem[{Kordopatis {et~al.}(2013)Kordopatis, Gilmore, Steinmetz, Boeche,
  Seabroke, Siebert, Zwitter, Binney, de~Laverny, Recio-Blanco, Williams,
  Piffl, Enke, Roeser, Bijaoui, Wyse, Freeman, Munari, Carrillo, Anguiano,
  Burton, Campbell, Cass, Fiegert, Hartley, Parker, Reid, Ritter, Russell,
  Stupar, Watson, Bienaym\'{e}, Bland-Hawthorn, Gerhard, Gibson, Grebel, Helmi,
  Navarro, Conrad, Famaey, Faure, Just, Kos, Matijevi\v{c}, McMillan, Minchev,
  Scholz, Sharma, Siviero, {Wylie de Boer}, \& \v{Z}erjal}]{Kordopatis2013}
Kordopatis, G., Gilmore, G., Steinmetz, M., {et~al.} 2013, AJ, 146, 134

\bibitem[{Lee {et~al.}(2011)Lee, Beers, An, Ivezi\'{c}, Just, Rockosi,
  Morrison, Johnson, Sch\"{o}nrich, Bird, Yanny, Harding, \&
  Rocha-Pinto}]{Lee2011}
Lee, Y.~S., Beers, T.~C., An, D., {et~al.} 2011, ApJ, 738, 187

\bibitem[{Leggett {et~al.}(1998)Leggett, Ruiz, \& Bergeron}]{Leggett1998}
Leggett, S.~K., Ruiz, M.~T., \& Bergeron, P. 1998, ApJ, 497, 294

\bibitem[{Loebman {et~al.}(2011)Loebman, Ro\v{s}kar, Debattista, Ivezi\'{c},
  Quinn, \& Wadsley}]{Loebman2011}
Loebman, S.~R., Ro\v{s}kar, R., Debattista, V.~P., {et~al.} 2011, ApJ, 737, 8

\bibitem[{{Martig} {et~al.}(2015){Martig}, {Rix}, {Aguirre}, {Hekker},
  {Mosser}, {Elsworth}, {Bovy}, {Stello}, {Anders}, {Garc{\'{\i}}a}, {Tayar},
  {Rodrigues}, {Basu}, {Carrera}, {Ceillier}, {Chaplin}, {Chiappini},
  {Frinchaboy}, {Garc{\'{\i}}a-Hern{\'a}ndez}, {Hearty}, {Holtzman}, {Johnson},
  {Majewski}, {Mathur}, {M{\'e}sz{\'a}ros}, {Miglio}, {Nidever}, {Pan},
  {Pinsonneault}, {Schiavon}, {Schneider}, {Serenelli}, {Shetrone}, \&
  {Zamora}}]{Martig2015}
{Martig}, M., {Rix}, H.-W., {Aguirre}, V.~S., {et~al.} 2015, MNRAS, 451, 2230

\bibitem[{Matteucci \& Francois(1989)}]{Matteucci1989}
Matteucci, F., \& Francois, P. 1989, MNRAS, 239, 885

\bibitem[{{Matteucci} \& {Greggio}(1986)}]{Matteucci1986}
{Matteucci}, F., \& {Greggio}, L. 1986, A \& A, 154, 279

\bibitem[{Minchev {et~al.}(2013)Minchev, Chiappini, \& Martig}]{Minchev2013}
Minchev, I., Chiappini, C., \& Martig, M. 2013, A \& A, 558

\bibitem[{Nidever {et~al.}(2014)Nidever, Bovy, Bird, Andrews, Hayden, Holtzman,
  Majewski, Smith, Robin, {Garc\'{\i}a P\'{e}rez}, Cunha, {Allende Prieto},
  Zasowski, Schiavon, Johnson, Weinberg, Feuillet, Schneider, Shetrone, Sobeck,
  Garc\'{\i}a-Hern\'{a}ndez, Zamora, Rix, Beers, Wilson, O'Connell, Minchev,
  Chiappini, Anders, Bizyaev, Brewington, Ebelke, Frinchaboy, Ge, Kinemuchi,
  Malanushenko, Malanushenko, Marchante, M\'{e}sz\'{a}ros, Oravetz, Pan,
  Simmons, \& Skrutskie}]{Nidever2014}
Nidever, D.~L., Bovy, J., Bird, J.~C., {et~al.} 2014, ApJ, 796, 38

\bibitem[{{Nidever} {et~al.}(2015){Nidever}, {Holtzman}, {Allende Prieto},
  {Beland}, {Bender}, {Bizyaev}, {Burton}, {Desphande}, {Fleming}, {Elia Garcia
  Perez}, {Hearty}, {Majewski}, {Meszaros}, {Muna}, {Nguyen}, {Schiavon},
  {Shetrone}, {Skrutskie}, {Sobeck}, \& {Wilson}}]{Nidever2015}
{Nidever}, D.~L., {Holtzman}, J.~A., {Allende Prieto}, C., {et~al.} 2015, ArXiv
  e-prints

\bibitem[{Prochaska {et~al.}(2000)Prochaska, Naumov, Carney, McWilliam, \&
  Wolfe}]{Prochaska2000a}
Prochaska, J.~X., Naumov, S.~O., Carney, B.~W., McWilliam, A., \& Wolfe, A.~M.
  2000, AJ, 120, 2513

\bibitem[{Ram\'{\i}rez {et~al.}(2013)Ram\'{\i}rez, {Allende Prieto}, \&
  Lambert}]{Ramirez2013}
Ram\'{\i}rez, I., {Allende Prieto}, C., \& Lambert, D.~L. 2013, ApJ, 764, 78

\bibitem[{{Randich} {et~al.}(2013){Randich}, {Gilmore}, \& {Gaia-ESO
  Consortium}}]{Randich2013}
{Randich}, S., {Gilmore}, G., \& {Gaia-ESO Consortium}. 2013, The Messenger,
  154, 47

\bibitem[{Reddy {et~al.}(2006)Reddy, Lambert, \& Prieto}]{Reddy2006}
Reddy, B.~E., Lambert, D.~L., \& Prieto, C.~a. 2006, MNRAS, 367, 1329

\bibitem[{Reid(2005)}]{Reid2005}
Reid, I.~N. 2005, ARAA, 43, 247

\bibitem[{Ro\v{s}kar {et~al.}(2008)Ro\v{s}kar, Debattista, Quinn, Stinson, \&
  Wadsley}]{Roskar2008}
Ro\v{s}kar, R., Debattista, V.~P., Quinn, T.~R., Stinson, G.~S., \& Wadsley, J.
  2008, ApJ, 684, L79

\bibitem[{Salaris {et~al.}(1993)Salaris, Chieffi, \& Straniero}]{Salaris1993}
Salaris, M., Chieffi, A., \& Straniero, O. 1993, ApJ, 414, 580

\bibitem[{Sch\"{o}nrich \& Binney(2009)}]{Schonrich2009a}
Sch\"{o}nrich, R., \& Binney, J. 2009, MNRAS, 399, 1145

\bibitem[{Sellwood \& Binney(2002)}]{Sellwood2002}
Sellwood, J.~a., \& Binney, J.~J. 2002, MNRAS, 336, 785

\bibitem[{Soderblom(2010)}]{Soderblom2010}
Soderblom, D.~R. 2010, ARAA, 48, 581

\bibitem[{Tinsley(1979)}]{Tinsley1979}
Tinsley, B.~M. 1979, ApJ, 229, 1046

\bibitem[{{van Leeuwen}(2007)}]{vanLeeuwen2007}
{van Leeuwen}, F., ed. 2007, Astrophysics and Space Science Library, Vol. 350,
  {Hipparcos, the New Reduction of the Raw Data}

\bibitem[{Wielen {et~al.}(1996)Wielen, Fuchs, \& Dettbarn}]{Wielen1996}
Wielen, R., Fuchs, B., \& Dettbarn, C. 1996, A \& A, 314, 438

\bibitem[{Yanny {et~al.}(2009)Yanny, Rockosi, Newberg, Knapp, Adelman-McCarthy,
  Alcorn, Allam, Prieto, An, Anderson, Anderson, Bailer-Jones, Bastian, Beers,
  Bell, Belokurov, Bizyaev, Blythe, Bochanski, Boroski, Brinchmann, Brinkmann,
  Brewington, Carey, Cudworth, Evans, Evans, Gates, G\"{a}nsicke, Gillespie,
  Gilmore, Gomez-Moran, Grebel, Greenwell, Gunn, Jordan, Jordan, Harding,
  Harris, Hendry, Holder, Ivans, Ivezi\v{c}, Jester, Johnson, Kent, Kleinman,
  Kniazev, Krzesinski, Kron, Kuropatkin, Lebedeva, Lee, Leger, L\'{e}pine,
  Levine, Lin, Long, Loomis, Lupton, Malanushenko, Malanushenko, Margon,
  Martinez-Delgado, McGehee, Monet, Morrison, Munn, Neilsen, Nitta, Norris,
  Oravetz, Owen, Padmanabhan, Pan, Peterson, Pier, Platson, Fiorentin,
  Richards, Rix, Schlegel, Schneider, Schreiber, Schwope, Sibley, Simmons,
  Snedden, Smith, Stark, Stauffer, Steinmetz, Stoughton, SubbaRao, Szalay,
  Szkody, Thakar, Thirupathi, Tucker, Uomoto, Berk, Vidrih, Wadadekar, Watters,
  Wilhelm, Wyse, Yarger, \& Zucker}]{Yanny2009}
Yanny, B., Rockosi, C., Newberg, H.~J., {et~al.} 2009, AJ, 137, 4377

\bibitem[{Zamora {et~al.}(2015)Zamora, Garc\'{\i}a-Hern\'{a}ndez, Prieto,
  Carrera, Koesterke, Edvardsson, Castelli, Plez, Bizyaev, Cunha, {Garc\'{\i}a
  P\'{e}rez}, Gustafsson, Holtzman, Lawler, Majewski, Manchado,
  M\'{e}sz\'{a}ros, Shane, Shetrone, Smith, \& Zasowski}]{Zamora2015}
Zamora, O., Garc\'{\i}a-Hern\'{a}ndez, D.~a., Prieto, C.~A., {et~al.} 2015, AJ,
  149, 181

\bibitem[{{Zucker} {et~al.}(2012){Zucker}, {de Silva}, {Freeman},
  {Bland-Hawthorn}, \& {Hermes Team}}]{Zucker2012}
{Zucker}, D.~B., {de Silva}, G., {Freeman}, K., {Bland-Hawthorn}, J., \&
  {Hermes Team}. 2012, in Astronomical Society of the Pacific Conference
  Series, Vol. 458, Galactic Archaeology: Near-Field Cosmology and the
  Formation of the Milky Way, ed. W.~{Aoki}, M.~{Ishigaki}, T.~{Suda},
  T.~{Tsujimoto}, \& N.~{Arimoto}, 421

\end{thebibliography}

\label{lastpage}

\end{document}